 \documentclass[preprint2]{aastex}

\newcommand{\myemail}{amorin@iaa.es}

\newcommand{\ha}{\relax \ifmmode {\mbox H}\alpha\else H$\alpha$\fi}
\newcommand{\hb}{\relax \ifmmode {\mbox H}\beta\else H$\beta$\fi}
\newcommand{\sii}{\relax \ifmmode {\mbox S\,{\scshape ii}}\else S\,{\scshape ii}\fi}
\newcommand{\nii}{\relax \ifmmode {\mbox N\,{\scshape ii}}\else N\,{\scshape ii}\fi}
\newcommand{\oii}{\relax \ifmmode {\mbox O\,{\scshape ii}}\else O\,{\scshape ii}\fi}
\newcommand{\oiii}{\relax \ifmmode {\mbox O\,{\scshape iii}}\else O\,{\scshape 
iii}\fi}

\slugcomment{To appear in {\it The Astrophysical Journal}}

\shorttitle{New insights on the ``Green Peas'' from OSIRIS-GTC}
\shortauthors{Amor\'in et al.}

\begin{document}

\title{The star formation history and metal content of the ``Green Peas''. 
New detailed GTC-OSIRIS spectrophotometry of three galaxies}

\author{R. Amor\'in \altaffilmark{1}, E. P\'erez-Montero and J.M. V\'ilchez}
\affil{Instituto de Astrof\'isica de Andaluc\'ia-CSIC, Glorieta de la Astronom\'ia S/N, E-18008 Granada, Spain}

\and

\author{P. Papaderos}
\affil{Centro de Astrof\'isica and Faculdade de Ci\^encias, Universidade 
do Porto, Rua das Estrelas, 4150-762 Porto, Portugal}


\altaffiltext{1}{CONSOLIDER-GTC fellow; \\ 
Email: \myemail}

\begin{abstract}
We present deep broad-band imaging and long-slit spectroscopy of three compact, 
low-mass starburst galaxies at redshift $z\sim$0.2--0.3, also 
referred to as Green Peas (GP).
We measure physical properties of the ionized gas and derive abundances 
for several species with high precision.
We find that the three GPs display relatively low
extinction, low oxygen abundances, and remarkably high N/O ratios
We also report on the detection of clear signatures of 
Wolf-Rayet (WR) stars in these galaxies.  
We carry out a pilot spectral synthesis study 
using a combination of both population and evolutionary
synthesis models. 
Their outputs are in qualitative agreement, strongly suggesting a 
formation history dominated by starbursts. 
In agreement with the presence of WR stars, these models show that these 
GPs currently undergo a major starburst producing between 
$\sim$4\% and $\sim$20\% of their stellar mass.
However, as models imply, they are old galaxies having had formed
most of their stellar mass several Gyr ago. 
The presence of old stars has been spectroscopically
verified in one of the galaxies by the detection of 
Mg{\sc i}$\lambda$$\lambda$5167, 5173 absorption line. 
 Additionally, we perform a surface photometry study based on 
HST data, that indicates that the three galaxies posses an exponential  
low-surface brightness envelope. If due to stellar emission, the
latter is structurally compatible to the evolved hosts of luminous 
BCD/H{\sc ii} galaxies, suggesting that GPs are identifiable
with major episodes in the assembly history of local BCDs.
  These conclusions highlight the importance of these objects 
as laboratories for studying galaxy evolution at late cosmic epochs. 
\end{abstract}

\keywords{galaxies: abundances --- galaxies: dwarf --- galaxies: evolution ---
 galaxies: starburst }

\section{INTRODUCTION}
\label{s1}

``Green Peas'' (GPs) are convenient laboratories to study galaxy
assembly at relatively low redshifts ($0.11 \la z \la 0.35$).  These
compact, high-surface brightness systems were recently identified on
images from the Sloan Digital Sky Survey (SDSS) Data Release~7 by
volunteers in the ``Galaxy Zoo'' project \citep{Lintott08,Lintott11}.
Their nickname reflects their point-like appearance and ``green''
color on SDSS image overlays.  The latter is a consequence of very
strong [\oiii]$\lambda$5007 line emission, with equivalent widths of
up to $\sim$2000 \AA, enhancing the observed SDSS $r$ fluxes at those
redshifts.  Such an extreme nebular emission contribution has so far
been documented in extremely metal-poor blue compact dwarf galaxies
(XBCDs) only \citep{Terlevich91,Izotov97,Papaderos98,Papaderos08} 
in addition to a few ultra-compact starbursting dwarfs in galaxy clusters
\citep{Reverte07}.  

The GPs were first studied in detail by \citet{Cardamone09} 
(hereafter C09), who
showed that these galaxies reside in lower-density environments and are
very rare ($\sim$2 galaxies deg$^{-2}$ brighter than 20.5 mag).  A
subset of 80 GPs with decent signal-to-noise (S/N) SDSS spectra was
spectroscopically characterized as purely starburst systems.  On
average, these galaxies appear luminous in both optical ($M_{B} \sim
-20$ mag) and UV ($L_{FUV} \sim 3\times$10$^{10} L_{\odot}$)
wavelengths, and are characterized by high surface brightness and 
very compact appearance (typical sizes $\la$5 kpc).  According to
C09, the GPs are low-mass galaxies (stellar masses M$_{\star}$ $<$
10$^{10.5}$M$_{\odot}$) with {prodigious} star formation rates (SFR up
to 60 M$_{\odot}$ yr$^{-1}$) and low intrinsic reddening 
(E(B-V)$\la$0.25). In particular, their specific star formation
rates (sSFR in the range 10$^{-7}$ to 10$^{-9}$ yr$^{-1}$) are among 
the highest inferred in the nearby Universe \citep[cf e.g.,
][]{Brinchmann04,Salim07}, and well in the range of those of
high-redshift galaxies \citep[e.g.,][]{Bauer05}.

Chemical abundances of the ionized gas have provided important
additional clues about the nature of the GPs.  
Oxygen abundance determinations, based on the direct ($T{\rm e}$) method 
led \citet{Amorin10}(hereafter A10) to conclude that the GPs in the 
Cardamone's sample are genuinely metal-poor galaxies, spanning a range 
of values $7.6\la$ 12$+$log(O/H) $\la 8.4$ with an average of one fifth 
of the solar value in their gas-phase metallicity in their gas-phase 
metallicity. Their results were recently confirmed by \citet{Izotov11} 
(hereafter Iz11), who pointed out that GPs are a subset of
luminous compact galaxies showing chemical abundances 
similar to lower-luminosity Blue Compact Dwarfs (BCDs).

Interestingly, the position of the GPs in the fundamental relation
between stellar mass and metallicity (the mass-metallicity relation,
MZR), and between $B-$band luminosity and metallicity (the
luminosity-metallicity relation, LZR) appear systematically offset 
(up to $\sim$0.3 dex in the MZR) to lower abundances when compared 
with the bulk of local star-forming galaxies (SFGs) from the SDSS 
\citep[A10, Iz11, see also][]{Amorin11}.  
Their location in the MZR, and also in the LZR, appear to form a
distinct sequence, along with nearby XBCDs
\citep[I Zw~18 or SBS~0335-052,] []{Guseva09} and some luminous
  BCDs at low \citep[][Iz11]{BergvallOstlin02} and intermediate redshifts
\citep[e.g.,][]{Hoyos05,Kakazu07,Salzer09} and most SFGs at high redshift 
\citep[e.g.,][]{Pettini01,Erb06,PM09,vanderWel11,Finkelstein11}.

In this context, ionic abundance ratios between species with an
assumed different stellar origin are important to probe the chemical
evolution of the GPs. This is the case of the nitrogen-to-oxygen ratio
(N/O), since the nitrogen and oxygen yields are driven by stars of 
different mass, therefore it gives relevant information about the SF 
rate and history of SFGs {\citep[e.g.][]{Molla06}}.  In metal-poor
SFGs (12$+$log(O/H)$\la$8), nitrogen production is expected to have 
mainly a primary origin, owing to massive stars. 
Then, in the O/H vs. N/O diagram, they form a {\it plateau} at 
log(N/O) $\sim$ -1.5 \citep{Alloin79,Campbell86,Izotov99,Pilyugin03}, 
with a relatively large vertical dispersion 
\citep[e.g.,][]{Garnett90,Pilyugin93,Henry06,Nava06,Vanzee06,PMC09}. 
For example, BCDs in the range 12$+$log(O/H)$\sim$7.6--8.2, 
generally show low N/O ratios between -1.54 and -1.27 \citep{Nava06}, 
with few exceptions only (see also Fig.~1 in \citet{Henry06}). 
By contrast, the nitrogen production in metal-rich SFGs 
(12$+$log(O/H)$>$8.2) has mainly a secondary origin,
powered by low- to intermediate-mass stars, which produce a positive 
correlation between N/O and O/H \citep[e.g.,][]{VilaCostas93}.
Intriguingly enough, {\em GPs show systematically larger N/O 
ratios compared to most SFGs at the same oxygen abundance}
\citep[A10, see also][]{Amorin11,Pilyugin12}, being in most cases 
located above the {\it plateau}.  
Their N/O ratios, however, seem to be normal for SFGs of similar 
stellar masses.

The known global properties of the GPs support the view that these 
galaxies go through a short and extreme phase in their evolution.
However, in order to confirm this and elaborate a coherent 
evolutionary picture for GPs, important pieces of the puzzle need to 
be supplied and investigated.  One example is the star formation 
history (SFH) of GPs, which still is not well-constrained, to a 
large extent because quantitative studies on the photometric 
structure and mass contribution of an underlying older stellar host 
are lacking.  In order to establish a physical and evolutionary 
connection between GPs and nearby BCDs, it is crucial to verify that 
the structural properties of the host galaxy in both SFG classes are 
compatible.  Whereas such a structural similarity has been 
demonstrated for luminous compact blue galaxies (LCBGs) at $z\sim 1$ 
by \citet{Noeske06}, no surface photometry studies for the lower-$z$ 
GPs exist as yet.  %
Another example is the interpretation of the relations between oxygen 
and nitrogen abundances, and with stellar mass and SFRs. 
They were discussed by A10 in terms of the balance
between inflows of metal-poor gas \citep[e.g.,][]{Koppen05} and the
presence of enriched outflows \citep[e.g.,][]{Vanzee98}.  However,
other scenarios usually invoked to explain large N/O ratios in nearby
metal-poor SFGs, such as possible pollution by Wolf-Rayet (WR) stars
\citep[e.g.,][]{Vanzee98,Brinchmann08,Monreal10,Berg11} were not addressed.

The main reason why the above questions remain open is that, until
now, GPs have been mainly studied using SDSS data.  This has 
allowed to infer relevant global properties for a large number of
galaxies. The downside, however, is that SDSS spectrophotometric 
studies are limited by the often poor signal-to-noise (S/N) and 
sensitivity of the data.  These limitations prevent, for example,
studies of faint spectral features, like those due to young WR 
stars or of weak stellar absorption features owing to an old 
underlying component. Moreover, the moderate to poor S/N of SDSS 
spectra impacts accurate chemical studies requiring precise 
measurements of, e.g., the [\oiii]$\lambda$4363\AA\ or 
[\nii]$\lambda$6584\AA\ line fluxes (e.g., A10) as well as stellar 
mass determinations based on a refined modeling of the spectral 
energy distribution (SED) of the stellar and gaseous continuum (Iz11).

In this study, we aim at going one step further in the understanding 
of the properties of GPs by using deep imaging and spectroscopy with 
the 10.4m Gran Telescopio Canarias (GTC), in addition to archival 
HST images.  These high-quality data allow us to analyze in better 
detail the chemical and structural properties of a small sample of GPs.  
In particular, deep spectra with the GTC are used both to improve on 
chemical abundance determinations and to study the SFH of GPs using 
population and evolutionary spectral synthesis models.  
Faint features, e.g., absorption lines from old stellar populations 
or signatures of massive and young WR stars, are examined.  
The latter will provide strong constraints on the age of the young 
starburst and on the amount of metal pollution in the interstellar 
medium that can be expected from them as well 
\citep[e.g., ][]{SchaererVacca98,Guseva00,Brinchmann08,PM11}.  

The paper is organized as follows: Section~\ref{s2} describes the
target selection, the GTC-OSIRIS observations and the data reduction,
and the HST data.  In Section~\ref{s3} we show the methodology and
present the results, which are then discussed in Section~\ref{s4}.
Finally, conclusions are given in Section~\ref{s5}. We assume a
standard cosmology with $H_{0} = 70, \Omega_{\rm \Lambda}=0.7$, and
$\Omega_m=0.3$.

\section{DATA}
\label{s2}

\subsection{Target selection}
\label{s2.1}

For the present study we have selected three GPs, 
\object[GP004054]{SDSS J004054.32+153409.6}, \object[GP113303]{SDSS 
J113303.79+651341.3}, and \object[GP232539]{SDSS 
J232539.22+004507.2} (for simplicity hereafter GP004054, GP113303,
and GP232539 respectively) from C09.  The main properties of the
galaxies as derived from the literature are summarized in 
Table~\ref{tbl-1}. The three galaxies are at very similar redshifts
$z \sim 0.24-0.28$ and they are located in relatively isolated
regions, with no nearby {\it bright} companions.  These GPs have a 
high UV surface brightness and luminosity ($\sim$10$^{10.5}
 L_{\odot}$ ), translating into large SFRs ($\ge 10 M_{\odot}$
yr$^{-1}$) per unit area. For this reason, they were also included 
in the sample of $\sim$30 super compact, UV luminous galaxies by
\citet{Hoopes07}, also known in subsequent studies as local
``Lyman-break Analogs''
\citep[LBAs][]{Basu07,Overzier08,Overzier09,Overzier10,Goncalves10}.
They were, therefore, better studied than other GPs. For example,
these galaxies are the three out of four GPs for which spatially
resolved HST imaging is publicly available up to date. From HST WFPC2
and ACS imaging \citep{Overzier09} we know that they share similar
morphologies, with UV and optical light dominated by few very luminous
star-forming clumps superimposed on a compact stellar host (optical
half-light radius of about 1 kpc).  

\subsection{Data set}
\label{s2.2}

Deep broad-band imaging and long-slit spectroscopy for the three
target galaxies were carried out using the OSIRIS instrument, mounted
on the 10.4~m GTC at the Observatory Roque de los Muchachos 
(La Palma, Spain).

OSIRIS\footnote{Detailed information on GTC and OSIRIS can be found in
  \url{http://www.gtc.iac.es}} \citep{Cepa2000} is an imager and
spectrograph for the optical wavelength range (from $\sim$3650 to 10000\AA),
located in the Nasmyth-B focus of GTC. It consists of two
2048$\times$4096 Marconi CCD42-82 with a 9.2 arcsec gap between them.
The unvignetted instrument field of view is 7.8$\times$7.8 arcmin with
a pixel scale of 0.125 arcsec.  Both imaging and spectroscopic
observations were obtained in service mode by the GTC staff during the
first semester of 2010. The log of observations is summarized in
Table~\ref{tbl-2}.

\subsubsection{GTC and HST Imaging}
 \label{s2.2.1}

Broad-band imaging for the three targets were obtained using the SDSS 
$z'$ filter (centered at 9695\AA) in the standard mode (2$\times$2
binning), given a pixel scale of 0.25 arcsec. In order to avoid 
substantial contamination from strong emission lines and reach faint 
surface brightness levels, we chose the $z'$ filter despite its
efficiency is lower than the $i'$ filter.  For each of the three
galaxies our broad-band images were taken under different sky
transparency conditions. 
Unfortunately, only GP113303 was observed in a dark night
under photometric conditions. The seeing was in all the cases below
1.2 arcsec. Total exposure times of 2250 sec were achieved taking
several series of five short exposures (90 sec) in a cross-shape
dithering pattern with offsets of 10 arcsec. Series of bias, twilight
sky flats, and several spectrophotometric standards  
were observed during the same nights (see \ref{s2.2.3}).

\begin{figure*}[ht]
\begin{center}
\includegraphics[angle=0,scale=0.70]{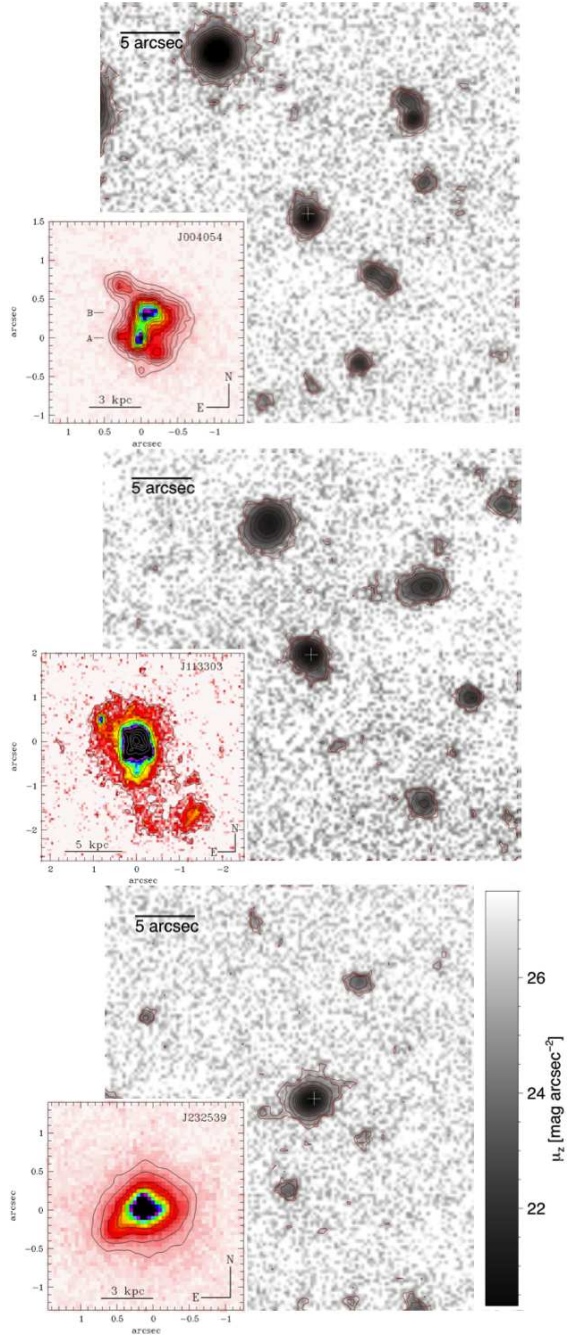}
\end{center}
\caption{Green Pea imagery: GTC-OSIRIS $z'-$band images of GP004054 (upper), 
GP\,113303 (middle), and GP\,232539 (bottom), are shown in surface 
brightness grey scale, as indicated by the bar in the lower right panel. 
In all cases contours indicate the $\mu_{\rm z} =$26 mag arcsec$^{-2}$ level. 
HST WFPC2 F606W ($R$) images, spatially resolving the central regions of 
these galaxies, are showed as insets in their lower left corner. 
North is up and east to the left.  }
\label{images}
\end{figure*}

The main goal of deep imagery with the GTC was the investigation 
of the close environment of our sample GPs e.g., extended tidal 
low-surface brightness features that could have gone undetected on 
the shallower SDSS and archival HST images.  
In Figure~\ref{images} we present deep $z'-$band images for the three 
GPs. The right gray scale bar and contours in these images show the 
$z'-$ band surface brightness of the galaxies.  
Only GP232539 appears to show a somewhat extended (size $>$ FWHM) 
LSB component.  
Possible small companions, barely seen on SDSS images, are 
projected few arc seconds to the galaxies.

In order to study the morphology and the structural properties of our 
sample galaxies, we additionally included in our analysis archival 
HST WFPC2 images in the filter F606W 
(HST proposal ID~11107, P.I: T. Heckman), which are also presented 
in Figure~\ref{images} as insets. 
These images reveal the complex morphology of the inner, high surface 
brightness regions of these GPs, and will be used for discussion in 
Section~\ref{s4.2}.
 
\subsubsection{GTC spectroscopy}
 \label{s2.2.2}

Long-slit spectroscopy was carried out in the standard mode
(2$\times$2 binning) with the highest resolution mode available at
these dates, R$\sim$1018 (at 5510\AA) and R$\sim$1122 (at 7510\AA).
Thus, we used the R1000B and R1000R grisms and a slit width of 0.8
arcsec, projecting onto a full width at half maximum (FWHM) of about 3
pixels. This setup yields wavelength coverages in the blue
$\sim$3630--7500\AA\ and in the red $\sim$5100--10250\AA, with typical
dispersion values of 2.1 and 2.6\AA\ pixel$^{-1}$, at 5510\AA\ and
7510\AA\ respectively. The spectra were taken {along} the parallactic
angle.  For each galaxy, blue and red spectra were taken during
different (dark) nights.  Seeing conditions varied between 1 and 1.4
arcsec, while atmospheric conditions were also not uniform, being two
of the nights spectroscopic, three of them clear, and the remaining
night slightly cloudy.  
Some Saharian dust was present in the atmosphere, affecting those 
observations at lower air masses significantly, reducing their final 
S/N, and doing more difficult the sky-background subtraction.
This is more evident in the red part of the spectra, were large sky 
residuals are present. 
Noisiest regions in the blue and red ends of the observed
spectral range were not considered in the subsequent analysis.  A series
of bias, twilight sky and dome flats, two calibration lamps, as well
as one or two spectrophotometric standards were observed during the
same nights.

\subsubsection{Data reduction}
\label{s2.2.3}

The data were fully reduced and calibrated using 
{\sc iraf}\footnote{{\sc iraf}: the Image Reduction and Analysis Facility 
is distributed by the National Optical Astronomy Observatories, 
which is operated by the Association of Universities for Research in 
astronomy, Inc. (AURA) under cooperative agreement with the National 
Science Fundation} routines.  This includes the usual procedures for
bias and overscan subtraction, flat-fielding corrections, cosmic ray
removal, and co-addition.  For the broad-band images, additional large
scale illumination and fringing patterns were removed 
effectively after subtraction of a ``master sky'' frame. 
This was obtained combining the 5x5 dithered science frames 
(after standard corrections and masking-out all saturated stars 
in the field) of each object. 
Finally we got deep ($\mu_{\rm z}$$\sim$26-27 mag arcsec$^{-2}$) images 
after a good sky-background subtraction.
For the spectra, wavelength calibration was done using HgAr+Xe+Ne lamp 
arcs. The accuracy ($\la$0.1\AA) was checked {\it a posteriori} using 
sky emission lines.
The spectra was corrected for atmospheric extinction and then flux
calibration was performed using several spectrophotometric standards
(Grw70$+$8247, Ross640, L1363-3, G157-34) along the observing blocks,
taken at the same night for each grating.

\section{GTC SPECTROSCOPY: ANALYSIS AND RESULTS}
\label{s3}

\begin{figure*}[ht]
\begin{center}
\includegraphics[angle=0,scale=0.85]{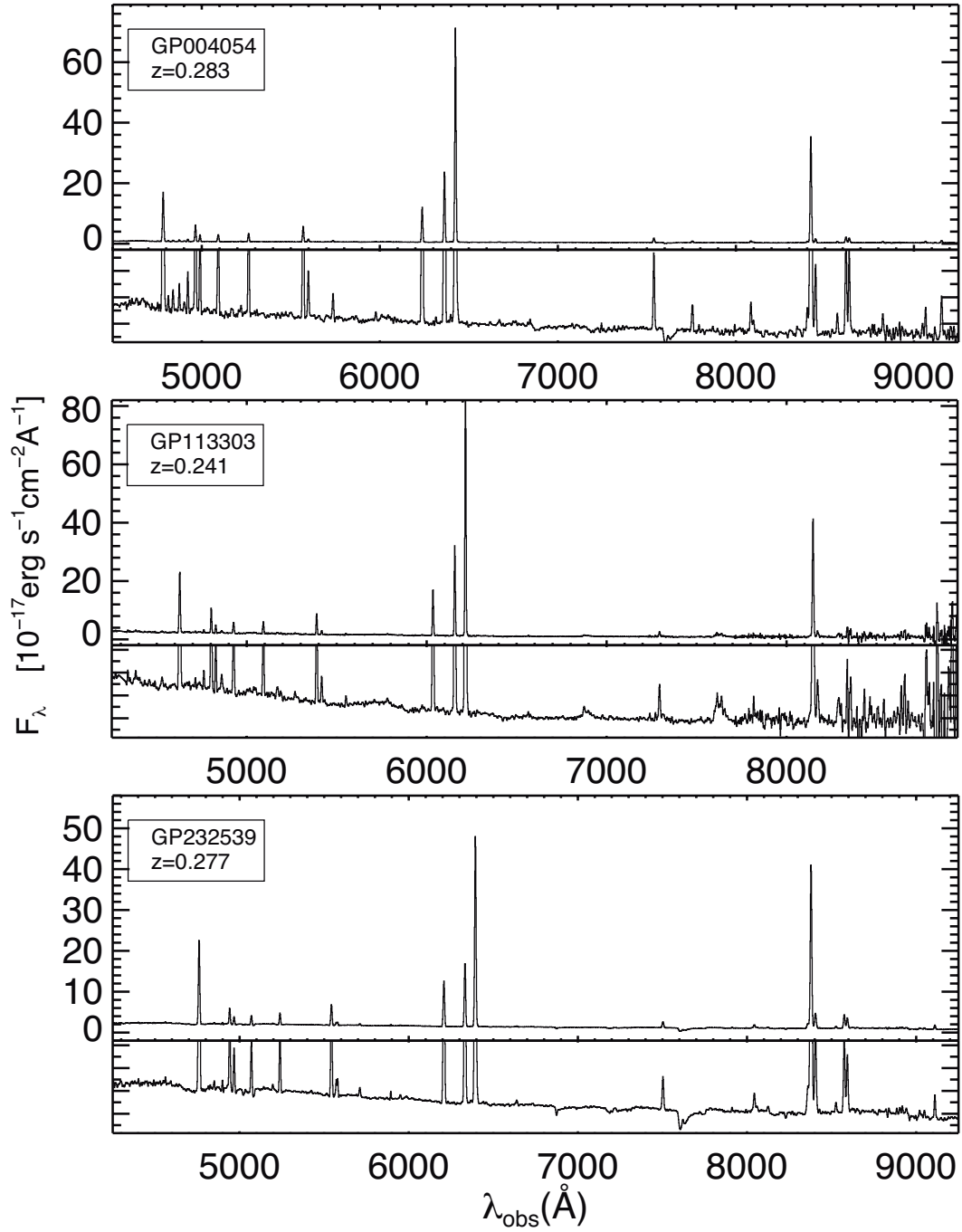}\\
\end{center}
\caption{The Deep GTC-OSIRIS spectra for the three GPs under study.
In order to highlight faint emission lines each spectrum is zoomed in 
the lower panel.}
\label{spectra}
\end{figure*}

In Figure~\ref{spectra} we present the OSIRIS-GTC spectra for the 
three GPs. The spectra are dominated by intense 
narrow nebular emission lines on top a faint blue stellar continuum 
lacking stellar Balmer absorption features.  The high S/N of spectra 
allowed for a detection of several faint emission lines (e.g. the 
temperature-sensitive [O{\sc iii}]4363 line) that are displayed 
magnified in the lower panel.

\subsection{Emission-line intensities and reddening correction}
 \label{s3.1}

We measured emission line intensities on the reduced and calibrated
spectra of the three galaxies using the task {\tt splot} of the
package {\sc iraf}. 
To measure the flux of a given line, we integrated the
flux between two points given by the position of a local continuum
placed by eye.  The statistical errors associated with the measured
emission line fluxes were calculated using the following expression
\citep{PMyD03}:

\[\sigma_i = \sigma_c \sqrt{N+\frac{W_i}{\Delta}} \]

\noindent where $\sigma_i$ is the error flux of the measured emission
line, $\sigma_c$ is the {\it rms} error derived for the local continuum,
$N$ is the number of pixels taken for the measurement of emission line
flux, $W_i$ is the absolute value of the emission line equivalent
width, and $\Delta$ is the wavelength dispersion.  Although there is
absorption of the Balmer emission lines caused by underlying stellar
populations in the objects \citep{Diaz88}, we checked in the
residuals to the {\sc starlight} fitting (see \S~\ref{s3.3}) that this 
effect is negligible compared to the reported errors.  
The emission line fluxes, F($\lambda$), relative to F(H$\beta$) = 1000, 
are listed with their corresponding errors in Table~\ref{tbl-3}.

Each emission line flux is affected by the presence of interstellar
dust which absorbs it according to the law:

\[ \frac{I(\lambda)}{I(H\beta)} = \frac{F(\lambda)}{F(H\beta)} 10^{-c(H\beta)f(\lambda)} \]

\noindent where I($\lambda$) and F($\lambda$) are the corrected and
measured emission line fluxes, respectively, c(H$\beta$) is the
constant of reddening, and f($\lambda$) is the extinction law, which
for this work, we took from \citet{Cardelli89}, and whose
values are also listed in Table~\ref{tbl-3} for the corresponding
listed emission lines. c(H$\beta$) was calculated for each object
as the error-weighted least square fit to the relation between the
extinction law and the quotient between the observed-to-theoretical
Balmer and H$\beta$ emission line fluxes 
before and after underlying continuum subtraction. 
We found that the differences between the corrected and uncorrected 
c(H$\beta$) values are always smaller than the uncertanties quoted 
in Table~\ref{tbl-3}. 
The Balmer emission-lines with enough S/N reach up to H17 for 
GP004054, H11 for GP113303, and H15 for GP232539. 
The theoretical emission-line ratios were derived
using \citet{Storey95} values for the appropriate electron
density and temperature for each one of the three galaxies.  Taking
the three objects coordinates into account to calculate the
corresponding Galactic extinction, which are c$_g$(H$\beta$) = 0.09
for GP004054, 0.02 for GP113303, and 0.05 for GP232539, we conclude
that the dominant contribution to reddening is due to intrinsic
extinction.  
The extinction-corrected emission line fluxes relative to I(H$\beta$) =
1000, along with their corresponding errors are listed in 
Table \ref{tbl-3}. 
In this table we also list the respective constants of reddening, the 
extinction corrected H$\beta$ flux and the H$\beta$ equivalent widths.

\subsection{Electron densities and temperatures}
 \label{s3.2}

All physical conditions, including electron density and temperature,
were calculated using the task {\tt temden} of the package {\sc
  iraf}, taking for each species the same atomic coefficients as in
\citet{Hagele06}.  All of them, as described in this subsection, are
listed in Table~\ref{tbl-4}.

Electron density has been estimated by taking the emission-line ratio
of [S{\sc ii}] 6716, 6731 {\AA}.  The error propagation from the
emission-line fluxes does not allow to give a precise estimate of the
density in any of the three galaxies, but it gives in all cases an
upper limit which is much lower than the critical density for
collisional deexcitation.

Electron temperature of [O{\sc iii}] was derived with high
precision in the three galaxies by taking the emission line ratio
between the sum of 4959, 5007 {\AA} and 4363 {\AA}.  This gives
temperatures for the three objects which are typical for BCDs and 
H{\sc ii} galaxies \citep{Campbell86,Masegosa94,Izotov06,Kehrig06,
Hagele06,Hagele08}, ranging from 13400 K for GP004054 and 14600 K 
for GP232539.

Other electron temperatures than [O{\sc iii}] were not derived 
directly in any of the three objects. Although [O{\sc ii}] 7319, 7330
{\AA}{\AA} and [N{\sc ii}] 5755 {\AA} were detected in the
spectra of the galaxies, they did not lead to a precise derivation of
the corresponding electron temperatures. Therefore, in order to
calculate the corresponding ionic abundances, we derived these
temperatures from t$_e$([O{\sc iii}]) and using the expressions
obtained from photoionization models described by \citet{PMyD03} and
\citet{PMC09} for t$_e$([O{\sc ii}] and t$_e$([N{\sc ii}]),
respectively.  Regarding t$_e$([S{\sc iii}]), as the nebular lines at
9069, 9532 {\AA} have not been detected, we derived this
temperature again from t$_e$([O{\sc iii}]), but using the empirical
relation obtained by \citet{Hagele06}. These derived temperatures are
also listed in Table~\ref{tbl-4}.

\subsection{Ionic and total chemical abundances}
 \label{s3.3}

Ionic abundances were calculated for the visible chemical species
in our optical spectra by using the task {\tt ionic} of the IRAF
package and taking the same atomic coefficients described in 
\citet{Hagele06} and the corresponding ionization correction factors
(ICF). The ionic and total abundances, along with these ICFs are
listed in Table~\ref{tbl-4}.

Helium abundances\footnote{No correction by neutral Helium nor by 
collisional excitation have been considered} have been obtained 
assuming that

\[\frac{He}{H} \approx \frac{He^++He^{2+}}{H^+} \]

He$^+$ abundances (noted as y$^+$ in Table~\ref{tbl-4}) were 
calculated as the error-weighted mean of the abundances derived using
the He{\sc i} emission lines at 4471, 5876, 6678 and 7065 {\AA}, with
t$_e$([O{\sc iii}]) and taking the expressions described by 
\citet{OliveySkill}. Only in the case of GP113303 we measured the
He{\sc ii} nebular emission line, but the corresponding y$^{2+}$
abundance is negligible as compared to y$^+$.
The derived He abundances (denoted as y in Table~\ref{tbl-4}) range 
from 0.087 to 0.089. These values fall between the pre-galactic He/H 
ratio of 0.08 \citep{Luridiana03} and the solar ratio 0.10 
\citep{Lodders03}, and are ``normal'' for  low-metallicity star-forming 
galaxies \citep[e.g., ][]{Kobulnicky96,Hagele08}. 

Oxygen abundances were calculated assuming that

\[\frac{O}{H} \approx \frac{O^++O^{2+}}{H^+} \]

\noindent with O$^+$ abundances derived from the [O{\sc ii}] 3727{\AA} 
emission line and taking t$_e$([O{\sc ii}]). In the case of O$^{2+}$, 
we toke [O{\sc iii}] 4959, 5007 {\AA} emission line 
intensities with t$_e$([O{\sc iii}]). 
The oxygen total abundances, ranging from 12$+$log(O/H) $=$ 7.81 for 
GP232539 to 7.98 for GP004054, are $\sim$ one fifth solar 
[12$+$log(O/H)$_{\odot} = 8.69$, \citet{Asplund09}]. 

Nitrogen abundances were calculated by assuming:

\[ \frac{N}{H} \approx ICF(N^+) \cdot \frac{N^+}{H^+} \]

\noindent and deriving N$^+$ abundances from the [N{\sc ii}] 6548,
6584 {\AA}{\AA} emission-line fluxes and t$_e$([N{\sc ii}]). The ICF
was calculated using the expression described in P\'erez-Montero
\& Contini (2009).  The derived N/O ratios, ranging from log(N/O) = -0.99
for GP232539 to -1.06 for GP004054, are well above the typical values  
corresponding to star-forming galaxies of the similar metallicity, 
i.e., the N/O {\it plateau} seen at log(N/O)$\sim -1.5$, 
\citep[e.g., ][]{Izotov99}.  

Neon abundances were calculated by assuming:

\[ \frac{Ne}{H} \approx ICF(Ne^{2+}) \cdot \frac{Ne^{2+}}{H^+} \]

\noindent calculating Ne$^{2+}$ ionic abundances from [{Ne{\sc iii}]
3868 {\AA} with t$_e$([O{\sc iii}]), and using the expression for the
corresponding ICF described by \citet{PM07}.  
The derived Ne/O ratios are all slightly higher than the solar value 
[log(Ne/O)$_{\odot} = -0.76$, \citet{Asplund09}].

Argon chemical abundances were calculated in GP004054, for which
both [Ar{\sc iii}] 7136 {\AA} and [Ar{\sc iv}] 4740 {\AA} were 
measured, and in GP232539, with the [Ar{\sc iii}] 7136 {\AA} emission 
line. Then, since we have both ionic argon abundance Ar$^{2+}$ and 
Ar$^{3+}$ in GP004054, and only the first one, Ar$^{2+}$, in GP232539, 
different ICFs were considered according to the expressions 
proposed by \citet{PM07}. 
As in the case of neon, both argon-to-oxygen ratios result higher 
than the solar value [log(Ar/O)$_{\odot} = -2.29$, \citet{Asplund09}].

Sulphur abundance was calculated by assuming the following:

\[\frac{S}{H} = ICF(S^++S^{2+}) \cdot \frac{S^++S^{2+}}{H^+} \]

\noindent with the S$^+$ derived from the [S{\sc ii}] 6717, 
6731{\AA}{\AA} and assuming that t$_e$([S{\sc ii}]) $\approx$ 
t$_e$([O{\sc ii}]).  In the case of S$^{2+}$, since our available 
spectral range has not allowed a measurement of the [S{\sc iii}] emission 
lines at 9069, 9532 {\AA}{\AA}, we derived their ionic abundances 
with [S{\sc iii}] 6312 {\AA}, and with t$_e$([S{\sc iii}]). 
We considered the ICF for (S$^+$+S$^{2+}$) obtained by \citet{PM06}. 
The derived sulfur-to-oxygen ratios range from -1.22 for GP113303 to 
-1.55 for GP232539. 
The high quoted errors for S/O make the derived values to 
be consistent with the solar value 
[log(S/O)$_{\odot} = -1.57$, \citet{Asplund09}].

Finally, iron abundance was calculated from Fe$^{2+}$ with the
emission line relative intensity of [Fe{\sc iii}] 4658 {\AA} and the
electron temperature of [O{\sc iii}]. We have also used the ICF
proposed by \citet{RyR04}. All Fe/O are similar in
the three galaxies, ranging from -1.49 for GP004054 to -1.69 for
GP232539. 

The derived Ne/O, S/O, Ar/O, and Fe/O values (see Table~\ref{tbl-4}) 
are, within uncertanties (which are especially high for S/O and Fe/O), 
consistent within the three galaxies. 
Overall, their values can be considered as ``normal'' when compared to 
the mean values found for nearby BCDs and H{\sc ii} galaxies in the 
literature \citep[e.g.,][]{Izotov99,Hagele06,Hagele08}.

\subsection{Spectral fitting}
 \label{s3.4}

In order to gain insights into the SFH of the GPs under study, we 
interpreted their integrated spectra by means of spectral synthesis 
models. 
To check the consistency of our results, we used both the population 
synthesis code {\sc starlight}\footnote{The {\sc starlight} project 
is supported by the Brazilian agencies CNPq, CAPES and FAPESP and by 
the France-Brazil CAPES/Cofecub programme. 
{\url http://www.starlight.ufsc.br}.}  
\citep{Cid04,Cid05,Mateus06} and a two-component evolutionary synthesis 
code that is based on} {\sc pegase 2.0} \citep{Fioc97}.

We used {\sc starlight} to synthesize the observed stellar continuum
of the galaxies as due to the superposition of single-age stellar
populations (SSPs) of different ages and metallicities.  
We used the SSP library provided by the 
{\sc popstar}\footnote{{\sc popstar} models are publicly available from 
{\url http://www.fractal-es.com/PopStar/SEDmod.html}} 
synthesis models \citep[hereafter {\sc run~1}, ][Garc\'ia-Vargas et al. 
in prep.]{Molla09,Manjon10}.  
These models follow the evolution of SSPs from very young (0.1 Myr) to 
very old (15.8 Gyr) ages by combining Padova '94 stellar evolution 
models with the most recent physics for stellar atmospheres and nebular 
continuum emission. {\sc popstar} models are therefore particularly 
well suited for the modeling of star-forming galaxies \citep{Molla09}. 
Note that the low spectral resolution (20 \AA) of the currently available 
{\sc popstar} SSPs does not permit a perfect match to Balmer stellar 
absorption features.  This, however, is not expected to have a notable 
impact on the derived SFHs.

In low-metallicity starburst galaxies, a significant contribution 
from nebular continuum emission is expected to be superimposed upon 
the stellar SED.  The gaseous emission not only enhances the 
luminosity of the galaxy but it additionally tends to make its 
spectral continuum redder than what is expected from a stellar SED 
\citep[e.g.,][Iz11]{Krueger95,Papaderos98}.  Therefore, as recently 
pointed out by Iz11, if nebular continuum emission is not taken into 
account in SED fitting for starburst galaxies the estimated stellar 
masses can be severely overestimated.  In this respect, the new 
{\sc popstar} SSPs including nebular continuum and line emission offer 
an important advantage towards a realistic SED modeling of GPs.

In modeling the stellar SED we used SSPs with three different 
metallicities, 0.008, 0.004, and 0.0004 (i.e., $\sim$1/2, 1/5, and 
1/45 solar) for a Kroupa initial mass function (IMF) between 0.15 
and 100 $M_{\odot}$, and all available SSPs between $\log t=$ 
5.0-10.2 yr.  For the SSPs with the youngest ages, i.e., $\leq$20 Myr, 
we only used models with $Z = 0.004$, which is the closest value to 
the gas-phase metallicity that we determined (see Table~\ref{tbl-3}) 
whereas for the older stellar component all three available 
metallicities were used. Prior to {\sc starlight} models, the flux 
calibrated spectra were de-redshifted and resampled to 1 \AA/pixel. 
Spectral regions with strong emission lines or sky-subtraction 
residuals were masked out from fitting. Models were applied on
the wavelength range 3500--6700\AA\ (rest frame) to exclude noisy 
spectral regions.  The S/N ratio of the spectra in the relatively 
featureless window between 4220 and 4280\AA\ varies from 15 to 45 
for the three galaxies.

\subsubsection{run 1: {\sc popstar} models}
\label{3.4.1}

\begin{figure*}[ht]
\includegraphics[angle=270,scale=.6]{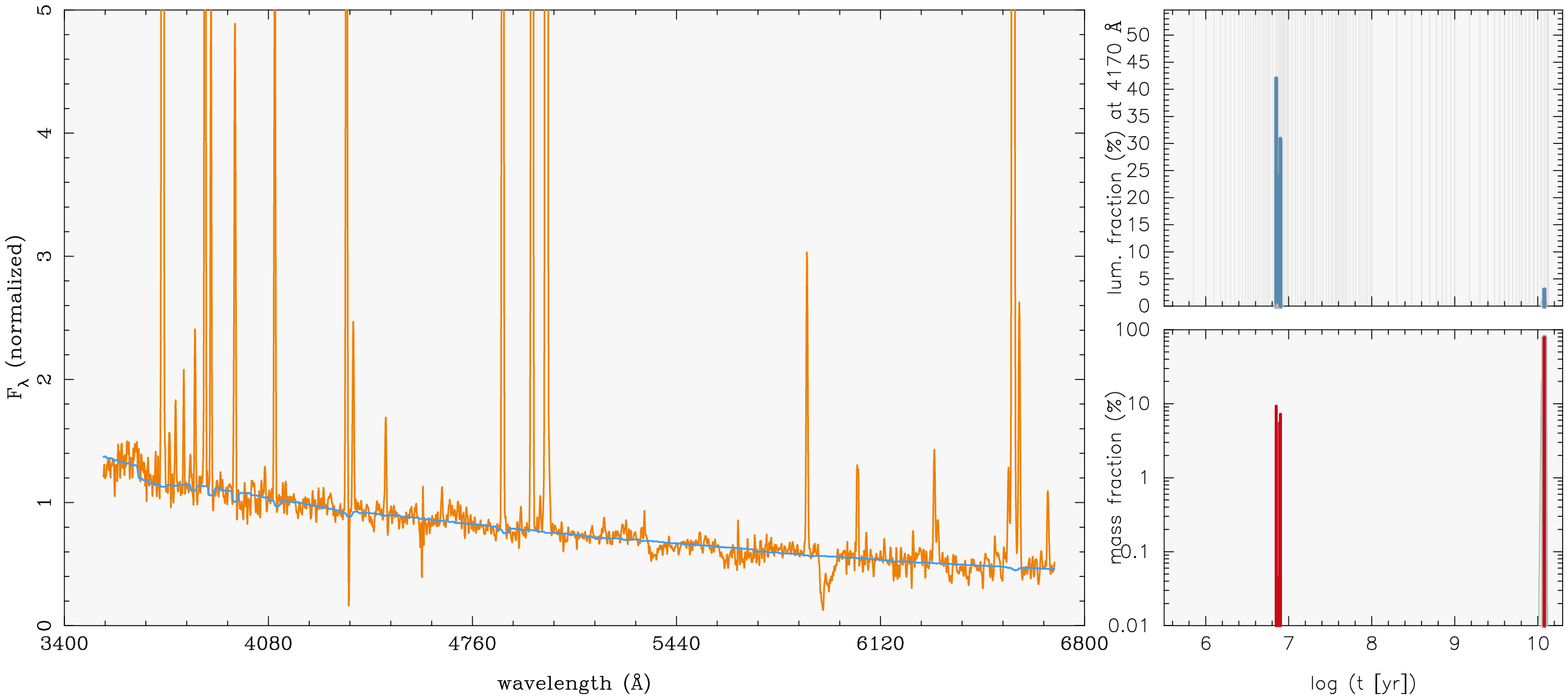} \\
\includegraphics[angle=270,scale=.6]{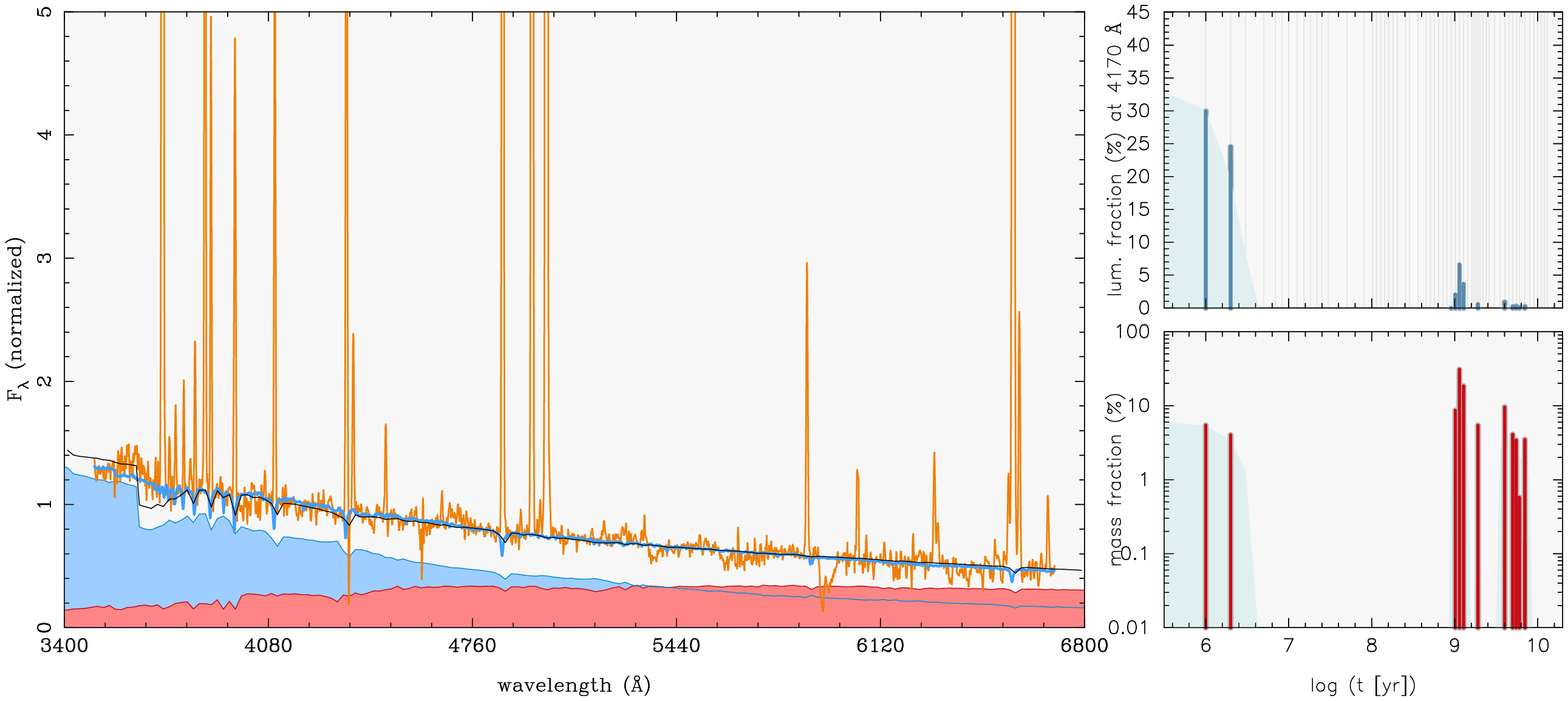} 
\caption{upper panel: Best-fitting synthetic SED based on 
{\sc popstar} SSPs ({\sc run1}; blue color), overimposed on the rest-frame 
observed spectrum of GP004054 (orange), normalized at 4170 $\AA$.
The smaller plots on the right show the luminosity and mass contribution of 
individual SSPs (upper and lower panel, respectively).
The age distribution of the library SSPs is illustrated by thin vertical 
lines in the upper panel.
lower panel: Fit to the observed spectrum based on purely stellar 
SSPs from Bruzual \& Charlot ({\sc run2}). The color coding is as in the 
upper large panel.
Vertical strips mark regions that have been flagged prior to spectral
fitting. 
The best-fitting synthetic SEDs from a two-component evolutionary 
synthesis model ({\sc run3}) that comprises an old and a young stellar 
component (red and blue shaded area, respectively), and self-consistently 
accounts for the observed Balmer H$\alpha$ and H$\beta$ EWs, is overlaid 
in black color.}
\label{SL1}
\end{figure*}

\begin{figure*}[ht]
\includegraphics[angle=270,scale=.6]{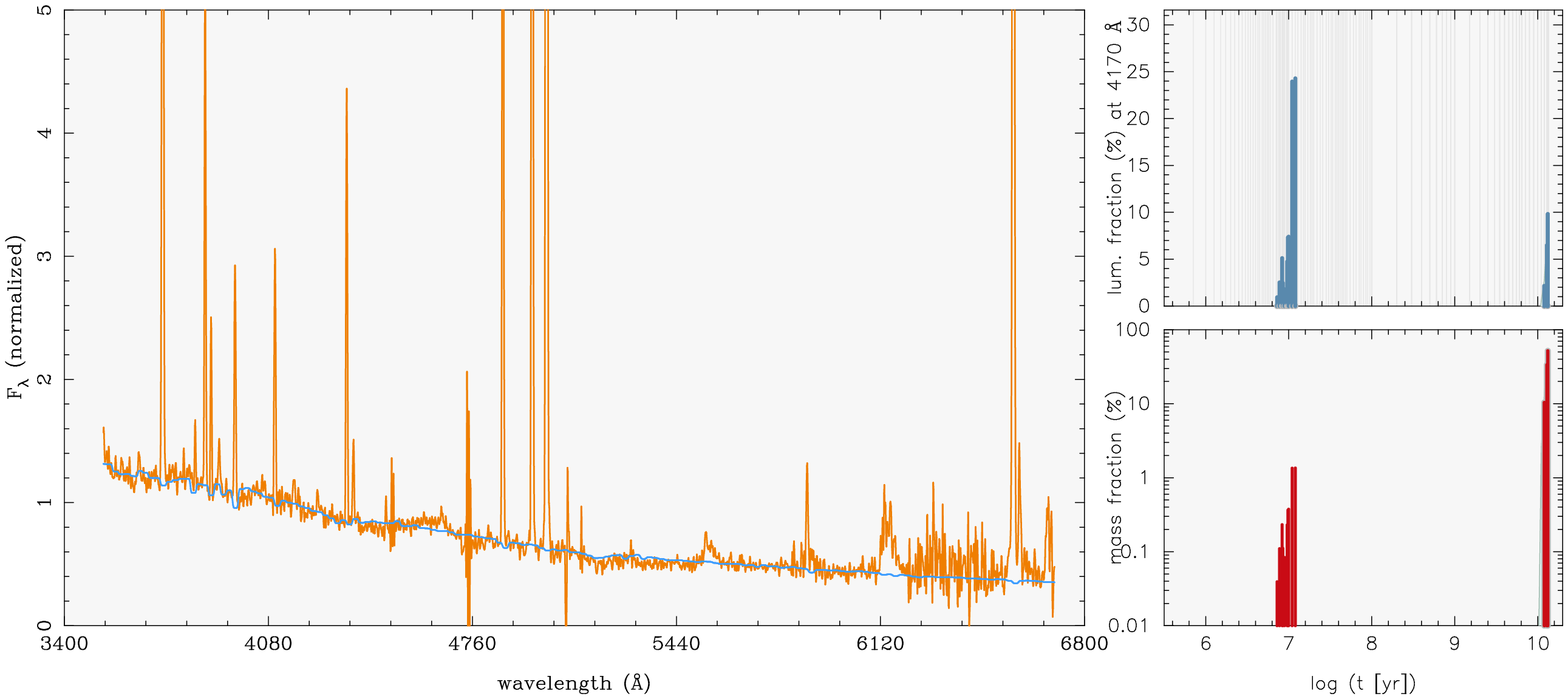}\\ 
\includegraphics[angle=270,scale=.6]{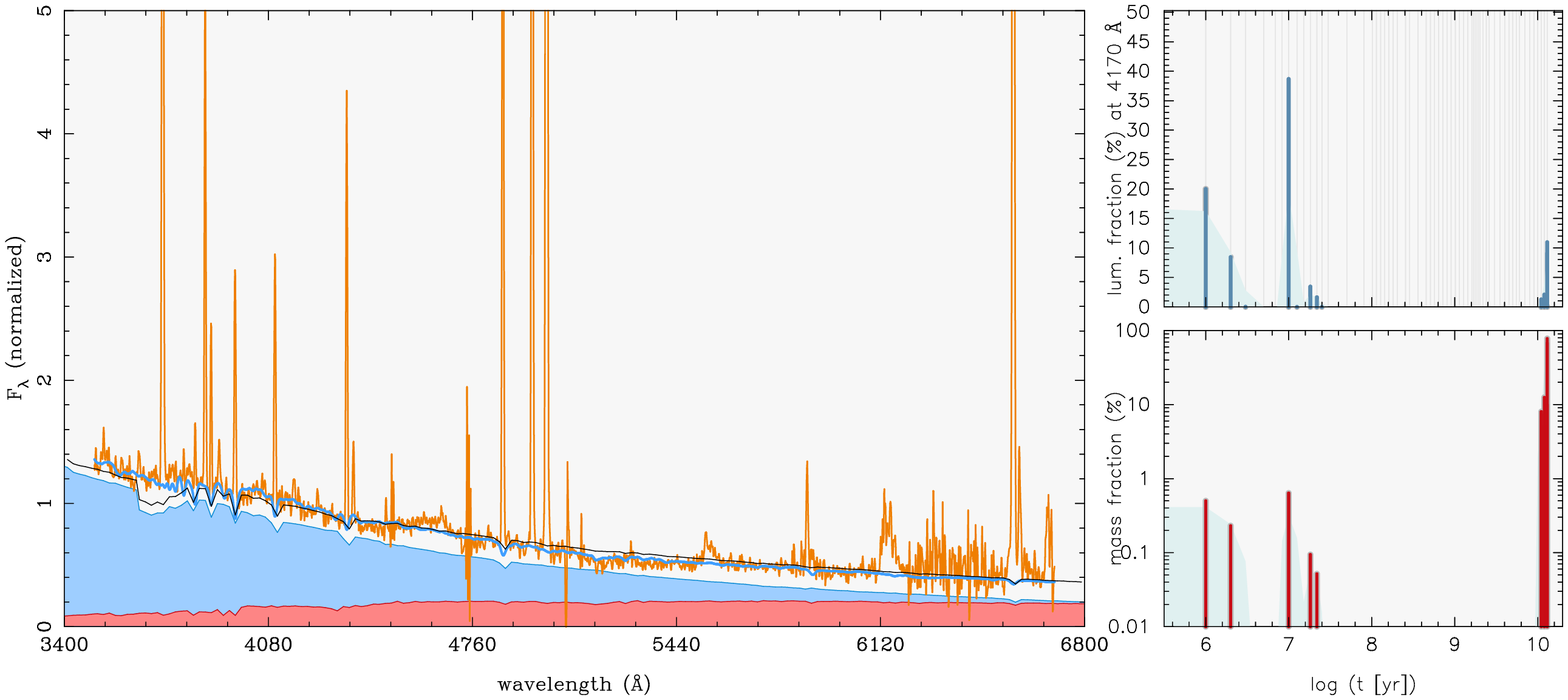} 
\caption{Same as Fig.\ref{SL1} for GP113303.}
\label{SL2}
\end{figure*}

\begin{figure*}[ht]
\includegraphics[angle=270,scale=.6]{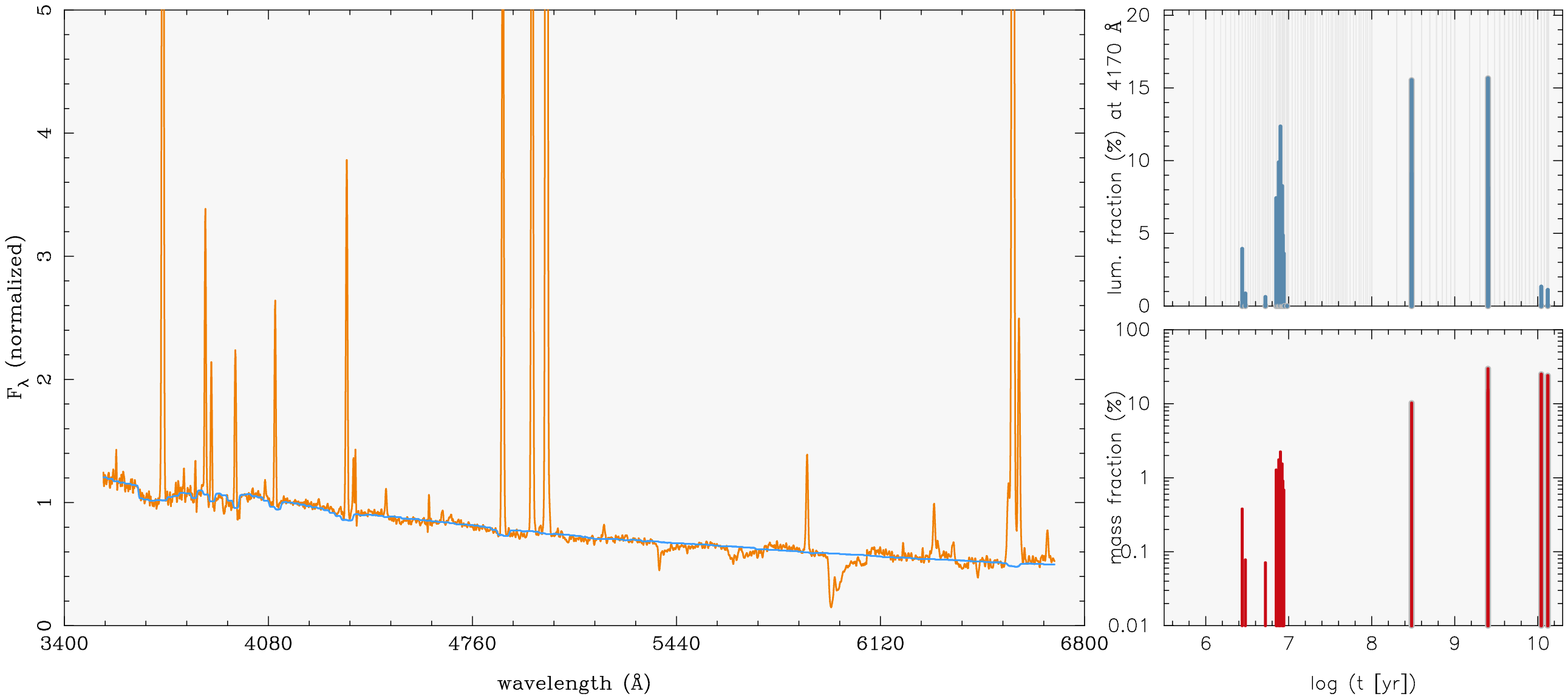}\\
\includegraphics[angle=270,scale=.6]{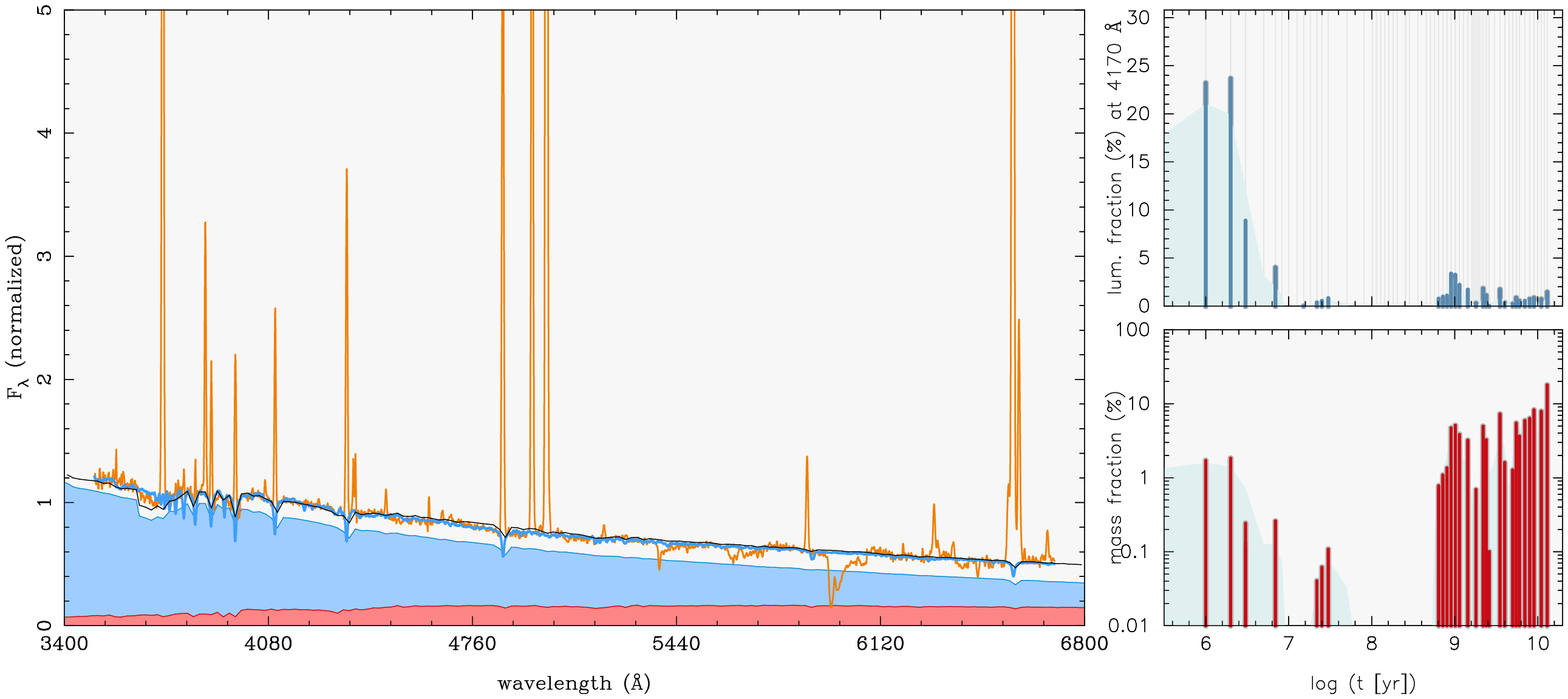}
\caption{Same as Fig.\ref{SL1} for GP232539. }
\label{SL3}
\end{figure*}

The {\sc starlight} fits based on the {\sc popstar} SSPs are shown 
overlaid with the observed spectra in the upper panels of 
Figures~\ref{SL1}-\ref{SL3}.  It can be seen that in all cases the 
synthetic stellar and gaseous SED provides a good match to the 
observations.
The right upper diagram shows the luminosity contribution (\%) of the
SSPs evaluated by {\sc starlight} to the normalization wavelength of
4170 \AA.  Vertical thin-gray lines depict the ages available in the
{\sc popstar} library. The stellar mass fraction corresponding to
each SSP is plotted on the lower panel.

The main output from the {\sc starlight} models for {\sc popstar} SSPs
is summarized in Table~\ref{tbl-5}.  Columns 2--4 list the predicted
relative contribution of the nebular emission to the total continuum 
emission at the [O{\sc ii}], H$\beta$ and H$\alpha$ wavelenghts.  
The obtained fraction of young ($\leq$100 Myr) stars 
$M_{\star, \rm young}$ 
with respect to the existing stellar mass of $\sim$20\%, suggests 
that GP004054 currently experiences a major episode in its assembly 
history.  The last two columns of Table~\ref{tbl-5} list the existing 
total stellar mass $M_{\star, \rm total}$ and the reduced $\chi^2$.  
It is interesting to point out that the stellar mass 
$M_{\star, \rm total}$ estimated from the fits, 
$\sim$2--3$\times 10^9$ $M_{\odot}$, places GPs in the range of
luminous BCDs \citep[e.g. ][]{BergvallOstlin02,Guzman03}, even when
aperture corrections (cf. Sect. \ref{s3.3.1}) are taken into account.
It can be seen that the {\sc starlight} solutions delineate two main 
SSP groups, a younger one dominating the optical light, and an older 
stellar component ($\sim$10 Gyr), contributing most ($\ga$80\%) of the 
stellar mass. Signatures for an intermediate-age stellar population 
were found for GP232539 only, where SSPs at $\sim$0.25 and $\sim$2.5 Gyr 
appear to make a substantial ($\ga 20$\%) contribution to the stellar mass.

Whereas the strong starburst activity in the GPs under study is
obvious, already from their high emission-line EWs, the inferred mass
fraction $M_{\star, \rm young}$ of stars recently produced is to be
considered with some caution.  Spectral synthesis models are known to
be plagued by substantial degeneracies, in particular for SFGs
\citep[see e.g. ][for a detailed discussion]{Guseva01} making
quantitative statements on the reality and relative importance of
individual features in the derived SFH difficult.  Additionally, it
should be called into attention that here we present a pilot attempt
of using {\sc popstar} SSPs in conjunction with {\sc starlight} and no
rigorous tests of possible numerical effects in this context have been
made so far.

We therefore include below two further modeling attempts in order to
gauge the variation of the derived $M_{\star, \rm young}$ for the GPs
under study. In the first one (hereafter {\sc run~2}), we 
employed {\sc starlight} using, however, SSPs based on the stellar 
models by \citet{B&C03}. 
One of the main differences between {\sc popstar} and 
Bruzual \& Charlot models is the inclusion of nebular continuum 
emission in the former (a detailed comparison of {\sc popstar} and other 
models can be found in \citet{Molla09}).
Additionally, the Bruzual \& Charlot SSPs have a higher spectral
resolution and assume a Salpeter IMF between 0.1 and 100 $M_{\odot}$.

\subsubsection{run 2: B\&C models}
\label{3.4.2}

For {\sc run~2} (lower panels of Figures~\ref{SL1}-\ref{SL3}) we
imposed roughly the same constraints as for {\sc run~1}, using the
same masks for emission lines (shaded areas) and normalization wavelength, 
and an SSP library with nearly the same age and metallicity coverage.
The main results as derived from {\sc run~2} are also included in 
Table~\ref{tbl-5}.  
Comparing them with those from {\sc run~1}, we found a satisfactory 
agreement in the overall SSP age distribution, with two prominent
peaks at young and old ages for GP113303 and signatures of an additional
intermediate-age stellar population in GP004054 and GP232539.  
In view of the differences in the SSP libraries and the details of 
the SFH obtained, some differences in, e.g., 
total stellar mass and young stellar mass fraction obtained from 
{\sc run} 1 and 2 are not surprising. 
With the exception of GP004054, {\sc run 1} yields a significantly lower 
total stellar mass and a higher mass fraction from young stellar populations
by a factor of about 2 ($\sim$0.3 dex) compared with those obtained
from {\sc run~2} (see Table~\ref{tbl-5}).

\subsubsection{run 3: evolutionary synthesis models}
\label{3.4.3}

As a second consistency check ({\sc run~3}), we applied a modified
version of the evolutionary synthesis code {\sc pegase 2.0}
\citep{Fioc97}.  In this approach, each spectrum was modeled as due to
the superposition of the SED from an old and a young stellar
population approximating, respectively, the underlying host galaxy and
the starburst component.  The host was modeled by an exponentially 
decreasing SFR since 13 Gyr with an e-folding time of 3 Gyr.
Note that the assumed SFH for the host implies for GP004050 
and GP113303 some contribution from intermediate-age to young 
(0.1--1 Gyr) stars which is not apparent from the {\sc starlight} 
fits ({\sc run1\&2}). 
As for the young stellar component, we assumed an instantaneous
burst whose age was allowed to vary between 0 and 50 Myr. 
The SEDs were computed assuming a fixed stellar metallicity of 
$Z_{\odot}/5$ and a Salpeter IMF (0.1--100 $M_{\odot}$), and include 
nebular continuum and line emission.

The best-fitting solutions were constrained by varying the burst age,
extinction C(H$\beta$) and bust parameter $b_{par}$ (mass fraction of
stars formed in the burst with respect to the mass of the stars ever
formed) that reproduce best the observed SED continuum and H$\alpha$
and H$\beta$ EWs. 
We note that the concept used here, i.e. the exploration of the SFH
and C(H$\beta$) that self-consistently account for Balmer line EWs, in
addition to stellar SEDs and colors of SFGs was originally used
for specific tasks in \citet{Izotov97} and \citet{Papaderos98} and
further developed in \citet{Guseva01} and \citet{Guseva07}.

In the lower panels of Figures~\ref{SL1}-\ref{SL3} we show the
best-fitting SEDs for the young and the old stellar component as blue
and red shaded areas, respectively.  It can be seen that the
superposition of a burst on an evolved galaxy host (black curve) 
can reproduce quite well the spectra of GP004054 and GP232539, whereas 
some systematic deviations are present in the case of GP113303.  
This might be attributed to the absence of an intermediate-age 
population in the latter system, as revealed by the {\sc starlight} 
models in {\sc run~1}\&{\sc 2}.  
The relevant model output is summarized in Table~\ref{tbl-6}.
Interestingly, in all cases the best-fitting models imply an
C(H$\beta$) very close to the measured value for the nebular
component. 
They also yield a good match to the observed H$\alpha$ and 
H$\beta$ EWs.
 
Tables~\ref{tbl-5} \& \ref{tbl-6} show the significant variation of
certain parameters (e.g. $M_{\star, \rm young}$) over {\sc run~1}
through {\sc run~3}.  While they may not be surprising in view of the
different models, SSP libraries and fitting constraints used, they
reflect the inherent uncertainties in the reconstruction of the SFH 
of individual starburst galaxies using state-of-art spectral synthesis 
models.
On the other hand, it is worth pointing out that all models, especially 
those including nebular emission ({\sc run~1} and {\sc run~3}), 
consistently suggest that our sample GPs presently undergo
a significant evolutionary stage in which they rapidly form between a
few \% and up to $\sim$20\% of their stellar mass. This highlights
the importance of these objects as laboratories for studying 
galaxy evolution at late cosmic epochs.

\subsubsection{Aperture corrections}
\label{s3.3.1}
 
We did not applied any aperture correction before fitting our spectra.
However, our flux calibration was checked using the SDSS spectra and
SDSS broad-band magnitudes. We calculated the fraction of total
optical light covered by the OSIRIS long-slit, deriving a slit
coverage a factor of 1.5--1.8 smaller than those of the SDSS fiber.
According to the fraction of light inside the SDSS fiber calculated in
A10, we estimated {a} slit coverage of about 45--55\% of the visible
light in the SDSS images.  Therefore, we must keep in mind that all
global properties that can be derived from the spectral analysis
(e.g., stellar mass) may be lower than their true values.

\subsection{Wolf-Rayet features}
 \label{s3.5}

All massive ($M \geq$ 25 $M_{\odot}$ for $Z_{\odot}$), very luminous
($10^5-10^6 L_{\odot}$) O stars pass through the WR phase 2$-$5 Myr
after their birth, spending less than $5\times10^5$ yr in this phase
{(e.g. Schaerer \& Vacca 1998, Meynet \& Maeder 2005)}.  The presence
of WR stars, as often observed in extragalactic H{\sc ii} regions
\citep{GonzalezDelgado94,GarciaVargas97}, and in both integrated
\citep[e.g.,][]{Allen76,Kunth81,Conti91,Masegosa91,Vacca92,Schaerer99,
  Guseva00, Brinchmann08} and spatially resolved
\citep{Kehrig08,Monreal10,LopezSanchez11} spectra of starburst galaxies,
is characterized by the detection of two broad emission features in
their optical spectra.  One of these features is the blue WR bump,
which is a blend of the N{\sc iii} $\lambda$4640, C{\sc iii}/C{\sc iv}
$\lambda$4650, and the broad He{\sc ii} $\lambda$4686 emission
line. This feature is mainly due to WN stars.  In contrast, WC stars
are the main responsible for the red WR bump, which is a blend of the
C{\sc iii} $\lambda$5698 and C{\sc iv} $\lambda$5808 broad emission
lines. The red WR bump is weaker than the blue one and
is rarely observed at low metallicities \citep{Crowther07}.  
WR stars are present in galaxies with recent starbursts 
\citep[e.g., BCDs, ][]{Guseva00,PM10}.
In GPs, however, no WR features had been detected so far.

\begin{figure}[ht]
\includegraphics[angle=0,scale=0.7]{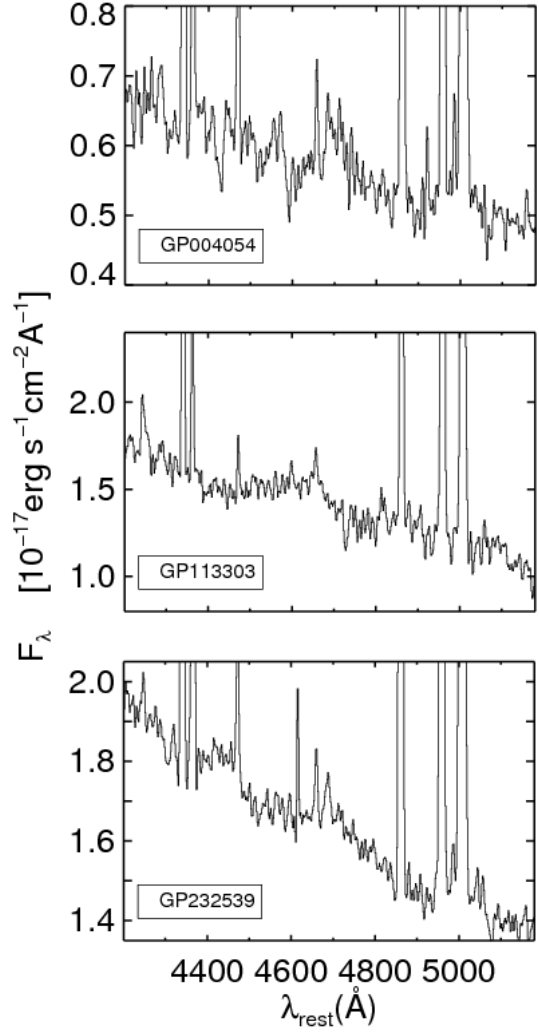}
\caption{Spectra of the galaxies zoomed in the region around the Wolf-Rayet 
blue bump.  }
\label{bumps}
\end{figure}

Figure~\ref{bumps} shows a zoom of the OSIRIS spectra in the
4200\AA--5200\AA\ (rest-frame) range, where the blue WR bump is
detected in the three GPs.  These WR features are also identified
after the subtraction of the continuum fitting by {\sc starlight}.  
In contrast, none of the galaxies has a red bump detection.

Given the low spectral resolution of our data, we measure the WR bumps
by using the {\sc iraf} task {\tt splot}.  First, we fitted and
substracted the adjacent continuum. Then we fitted gaussians to the
nebular emission lines in the region between 4600 and 4750\AA, and
finally we integrated the remaining emission above zero to have an
estimate of the blue WR bump fluxes. The main properties derived from
these measurements are summarized in Table~\ref{tbl-6}. For example,
from the luminosity of the bump, and adopting a theoretical WR
luminosity from {\sc starburst99} models \citep{Leitherer99} at the
same metallicity, we inferred the number of WR stars expected for each
galaxy.

\section{DISCUSSION}
\label{s4}

\subsection{Stellar mass}
 \label{s4.1}

In Table~\ref{tbl-5} we list stellar masses contained in the
spectroscopic aperture as derived from the spectral fitting, and
using the adopted distances to the galaxies (see Table~\ref{tbl-1}),
and mass-to-light ratios obtained from the mixture of {\sc popstar} 
synthetic stellar populations fitted to each galaxy spectrum.  
We also give in Table~\ref{tbl-5} an estimate of the percentage of 
the stellar mass given by stellar populations younger than 20 Myr, 
and the fractional contribution --- for three different spectral 
ranges --- of the nebular continuum.
 
Our stellar mass estimates from {\sc run~1} 
for GP004054 ($M_{\star} =$1.7$\times$10$^{9} M_{\odot}$), 
GP113303 ($M_{\star} =$2.8$\times$10$^{9} M_{\odot}$), and GP232539 
($M_{\star} =$2.4$\times$10$^{9} M_{\odot}$), 
confirm that these galaxies are {\it dwarfish systems} having had 
formed between a few \% and up to $\sim$20\% of their present stellar 
mass in a recent ($\leq$100 Myr) major star formation episode.
Note that the stellar masses derived here are not far from those
inferred for the same objects in previous studies (see Table~\ref{tbl-1}),
despite significantly different modeling approaches used.  

Discrepancies are likely due to the inclusion of the contribution 
of the nebular continuum emission ({\sc run~1}) in this study.
For the three analyzed objects, we find this contribution 
to be significant and to increase to several \% towards the red spectral 
range (see Table~\ref{tbl-5}).
The impact of the red nebular continuum on SED fitting is already
suggested by the fact that {\sc starlight} models based on purely
stellar SSPs ({\sc run~2}) yield a by a factor 
$\sim$2 lower $M_{\star, \rm young}$, consequently a larger mass 
fraction of old stars.  
Precisely this effect was discussed by Iz11 and invoked
in order to explain the differences between their $M_{\star}$ values
and those derived by C09 for the whole sample of GPs.   
Note, however, that we do not find any clear correlation between the
offset of the derived stellar masses and the relative flux
contribution of the nebular continuum in our three studied objects.

For GP004054, C09 and Iz11 did not published $M_{\star}$ values.
However, we compared our estimates with those given by
\citet{Overzier09}.  The latter were obtained from the SDSS/DR7
stellar mass catalog
\footnote{Available at {\url
    http://www.mpa-garching.mpg.de/SDSS/DR7/.}}, and derived using
SDSS photometry and a large grid of models constructed on the basis of
the \citet{B&C03} library.  Interestingly, stellar masses calculated
here for GP004054, and also for GP113303 and GP232539 (also listed by
\citet{Overzier09}) are within $\sim$0.3 dex in agreement with our
values.
It is worth noting that values given by \citet{Overzier09} did not take 
into account any contribution from nebular continuum, which may be 
possibly masked or compensated by other uncertainties.

\subsection{The star formation history of the GPs}
 \label{s4.2}
\begin{figure}[ht]
\begin{center}
\includegraphics[angle=0,scale=.80]{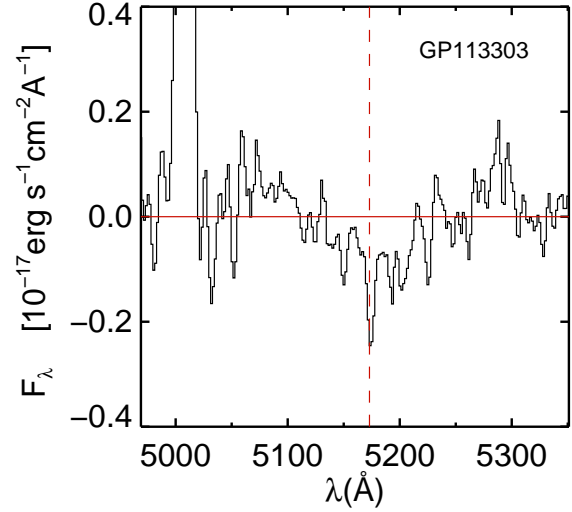}
\caption{Spectrum of GP113303 zoomed in the region around the Mg{\sc i}
  absorption line after continuum subtraction. The central
wavelength of the Mg{\sc i}  $\lambda$5173 is indicated by the
  vertical dashed line.}\label{fig-Hg}
\end{center}
\end{figure}

Several constraints on the SFH of the studied GPs can be inferred
from our results.  
As consistently indicated by spectral synthesis models, none of them 
presently forms its first stellar population. Quite to the contrary, 
an evolved stellar component with an age between a few $10^8$ yr and 
several Gyr is present in all three GPs and provides at least $\sim$80\% 
of their stellar mass. 
In particular, direct evidence for an old stellar component is found  
for the system GP113303, where we detect a faint ($EW \sim 1.5$\AA) 
but clear Mg{\sc i}$\lambda$$\lambda$5167, 5173 broad absorption 
line.  This feature is due to the presence of old stars and, as 
shown in Fig~\ref{fig-Hg}, it is clearly visible by eye in the 
spectra.  At least for GP113303, the detection of this feature gives 
us the confirmation of the presence of old stars.  Whereas for 
GP004054 and GP232539 the spectral fitting led to similar results 
than for GP113303, absorption features in their spectra are probably 
too faint to be detected.

Both population and evolutionary synthesis models including the 
nebular component ({\sc run 1\&3})
imply, on the other hand, that the GPs studied here are experiencing 
a significant stage in their buildup, as they form $\sim$4--20\% of 
their total mass in an intense intermittent or prolonged starburst 
episode which, as suggested by the presence of WR features is very 
recent or still ongoing.

The strong starburst (SFR$\sim 7-14 M_{\odot}$ yr$^{-1}$, see 
Table~\ref{tbl-1}) taking place in these galaxies and their low 
stellar masses ($\sim 1.5-2.7 \times$10$^{9} M_{\odot}$ yr$^{-1}$) 
imply large {\it specific} star formation rates, 
sSFR$\sim 2-9\times$10$^{-9}$ yr$^{-1}$. 
In agreement with C09 and Iz11, these values are typical for GPs  
but they are unusually high in the local Universe \citep{Brinchmann04}, 
being more comparable to those observed in high redshift starburst 
galaxies. Since these sSFRs imply relatively short mass doubling times 
($\equiv$ 1/sSFR) $\la$500 Myr), it is likely that the GP evolutionary 
stage is very brief. 
This is also to be expected from the enormous energetic release 
from the starburst that will quickly heat up and disperse the cold gas
reservoir of GPs, quenching star-forming activities after a few Myr.
The starburst nature of GP is also supported by {\sc starlight} fits 
which suggest longer quiescent phases preceeding the current burst
ranging from a few $10^8$ yr in the case of GP232539 to
$\sim$1--10 Gyr for GP004054 and GP113303.
Note that the smoother SFH of GP232539, resulting into a substantial 
intermediate-age stellar component is indicated by both {\sc run~1} and 
{\sc run~2}.

The overall evidence (stellar mass, SFH pattern and burst parameter) is
consistent with the hypothesis of GPs being identifiable with previous 
starburst-dominated phases in 
the lifetime of luminous BCDs in the local Universe.
As apparent from HST/WFPC2 $R$-band (F606W) images (Fig.~\ref{images}), 
the GPs under study are as well compatible to local BCDs with respect 
to their compactness ($\sim$5 kpc) and the irregular morphology of their 
high-surface brightness component. 
Furthermore, similar to what is found for the main population of 
chemically evolved (7.6$\la$12+(O/H)$\la$8.3) nearby BCDs 
\citep[e.g.,][]{Papaderos96,Cairos01,Cairos03,Noeske03,GdP05,Vaduvescu06,
Corbin06,Amorin07,Amorin09}, the GPs studied here show a more extended 
lower-surface brightness (LSB) envelope that is presumably due to an 
old stellar host. 
It is worth pointing out, on the other hand, that, contrary to 
most BCDs, the LSB host of GPs displays significant departures 
from ellipticity, suggesting a lesser degree of dynamical relaxation,
or merger origin, which appears rather typical for the high-luminosity end
of local BCDs \citep[e.g.][]{TMT97,Ostlin01,BergvallOstlin02,Adamo11}
This is particularly true for GP113303 for which HST data reveal 
significant LSB emission out to $\sim$10 kpc southwest of its nuclear 
region.

\begin{figure}[ht]
\begin{center}
\includegraphics[angle=270,scale=0.45]{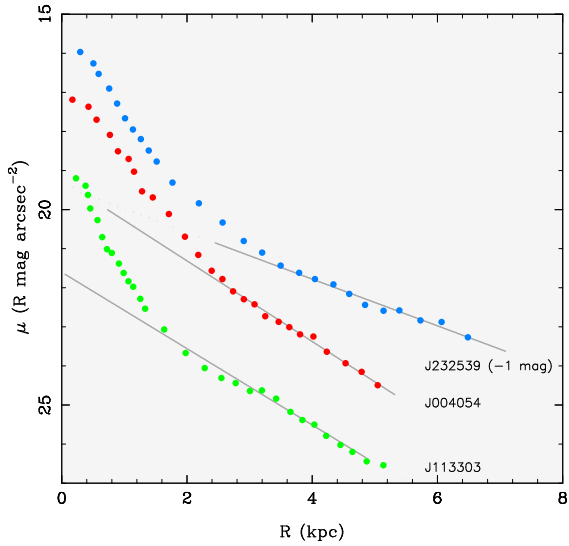}
\end{center}
\caption{Surface brightness profiles (SBPs) of the three GPs derived 
from archival HST/WFPC2 $R-$ band (F606W) images. 
The SBPs have been corrected for Galactic
extinction and cosmological dimming but no $k$ corrections were applied.
The SBP of J232539 has been shifted by --1 mag for the sake of clarity.
It can be seen that all SBPs show an extended exponential LSB envelope
dominating for $\mu \ga 22-23$ mag~arcsec$^{-2}$. Linear fits to the LSB
component (gray lines) yield scale lengths and central surface
brightness levels of 1.5--3 kpc and 19--21.5 mag~arcsec$^{-2}$, 
respectively.
}
\label{SBPs}
\end{figure}

\begin{figure}[ht]
\begin{center}
\includegraphics[angle=270,scale=0.35]{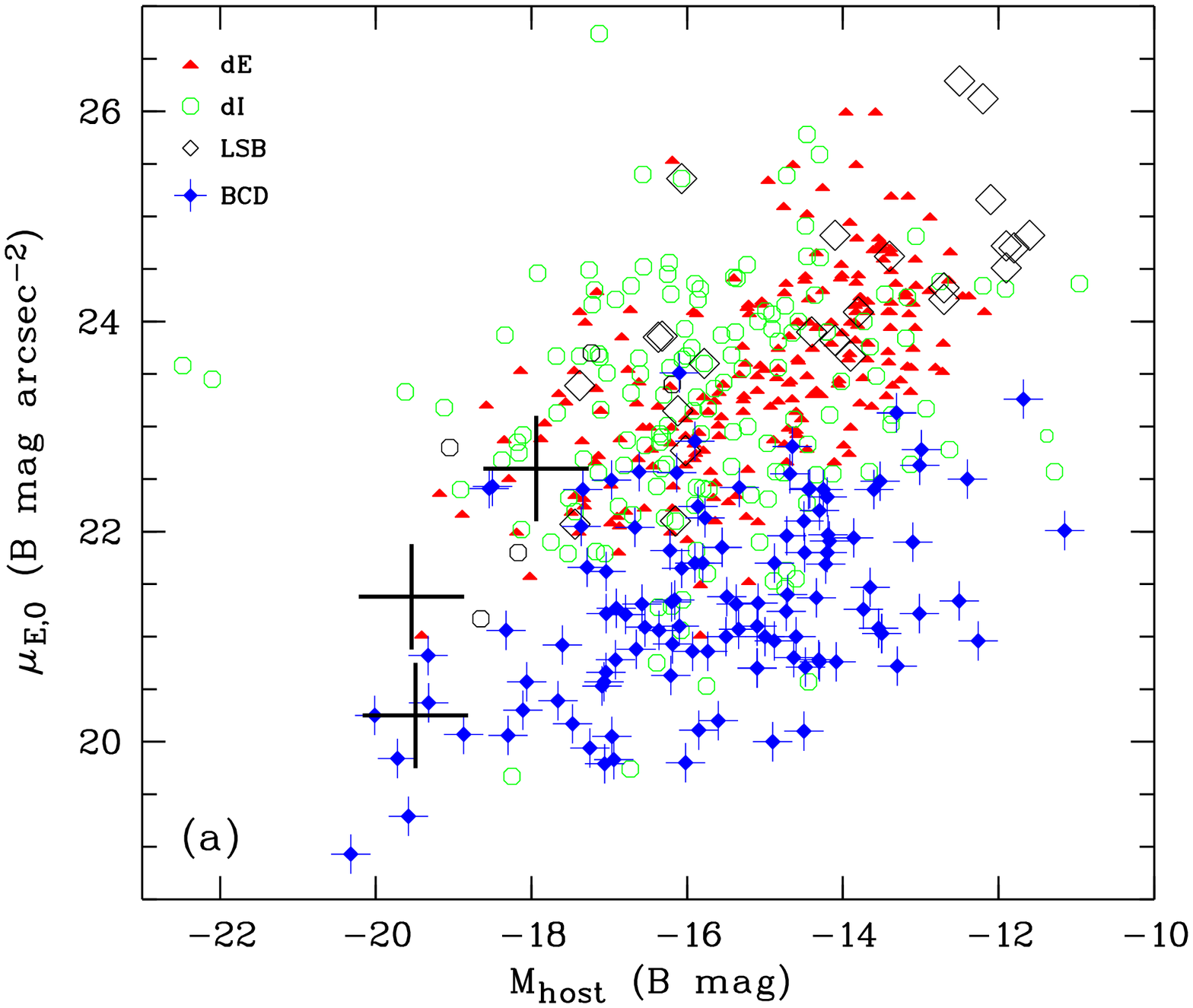}\\
\includegraphics[angle=270,scale=0.35]{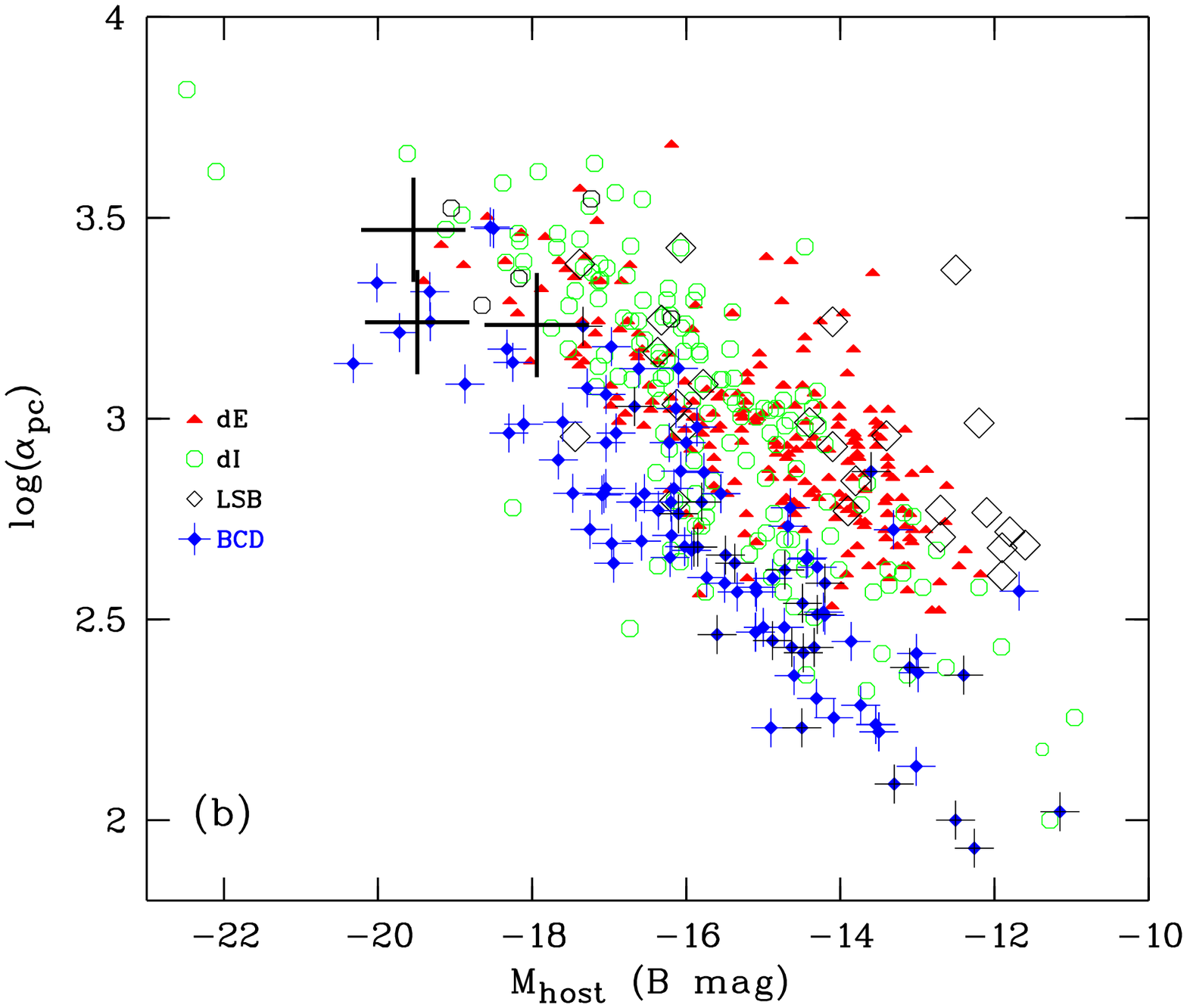}
\end{center}
\caption{Comparison of the structural properties of the LSB component of our
sample GPs (large crosses) with those of other types of low-mass galaxies
compiled in \cite{Papaderos08}. 
Photometric quantities derived from fitting of $R$ band profiles 
in Fig.~\ref{SBPs} have been converted into $B$ band assuming a $B-R=1$ mag. 
The lower and upper panel compare, respectively, the central surface
brightness and the exponential scale length vs the absolute magnitude 
of the LSB component of GPs with those of 
BCDs, dwarf irregulars (dIs), dwarf ellipticals (dEs) and LSB galaxies. 
It can be seen that GPs occupy roughly the same parameter space as luminous
BCDs.
 }
\label{host}
\end{figure}

The properties of the LSB component can be better quantified from the 
surface brightness profiles (SBPs) in Fig.~\ref{SBPs}. 
The SBPs were derived using method~iv in \citet{Papaderos02} and were 
converted to the Vega system, following the prescriptions by 
\citet{Holtzman95}. They have been corrected for cosmological dimming as 
$\mu_{\rm corrected} = \mu_{\rm observed}-10\,\log(1+z)$ and for
Galactic extinction but no $k$ corrections were applied.
It can be seen that all SBPs show an outer exponential intensity 
decrease in their LSB envelope, as is typically the case for nearby BCDs.
On the assumption that the LSB emission is of stellar origin, one can 
infer from linear fits to the exponential part of the SBPs 
the scale length and central surface brightness of GP hosts to 
1.5--3 kpc and 19--21.5 mag~arcsec$^{-2}$, respectively.
In Fig.~\ref{host} we compare the structural properties of the host
galaxy of GPs with those of local BCDs, dwarf irregulars (dIs), LSB 
galaxies and dwarf ellipticals (dEs). 
It can be seen that GPs fall in the parameter space that is populated 
by luminous BCDs, indicating that these two galaxy classes are very 
similar structurally.

One word of caution is in order here. A roughly exponential intensity 
drop off is a generic property of extended nebular halos that are
expected to arise in galaxies with strong starburst activity 
\citep{Papaderos02}.
 An exponential profile in the LSB periphery of GPs should not
therefore be taken as a foolproof sign of an extended \emph{stellar}
host galaxy. If the LSB envelope is partly or entirely due to extended
nebular emission, then the absolute magnitude of the GP host will
decrease by 0.75--1 mag, with a simultaneous brightening of the
central surface brightness and decrease of the exponential scale
length. This would move the GPs under study downstream in either plot,
in better agreement to lower-luminosity BCDs.

\subsection{The nitrogen over-abundance in GPs}
 \label{s4.3}

The good quality of the GTC data allowed {us to derive the} physical 
properties and chemical abundances of the ionized gas in the sample GPs 
with much better accuracy than previously done for these galaxies. 

{The derived} electron temperatures and densities (T$_{e}$([\oiii]) $\sim$ 
13000--15000 K, n$_{e}$([\sii]) $<$ 600 cm$^{-3}$) are consistent to those 
measured by A10 and by Iz11 for a larger sample of GPs. 
Moreover, these values are in general agreement with those typically 
measured in other samples of nearby star-forming dwarf galaxies 
\citep[e.g.,][]{Campbell86,Masegosa94,Lee04,Hagele08}. 

The oxygen abundances (12$+$log(O/H)) found for the sample GPs are 
7.81$\pm$0.14, 7.91$\pm$0.10, and 7.98$\pm$0.06 for GP232539, GP113303, 
{and} GP004054 respectively.  
These metallicity values are also consistent with those previously 
published using SDSS data (see Table~\ref{tbl-1}), and 
confirm that the three GPs {\it are metal-poor galaxies}.

On the other hand, the three GP galaxies appear to be relatively 
{\it rich in nitrogen}.
The logarithmic N/O ratios {determined} for the sample GPs are 
--1.06 $\pm$ 0.04, --1.04 $\pm$ 0.08, and --0.99 $\pm$ 0.11 for 
GP004054, GP113303, and GP232539, respectively. 
These values are slightly higher (0.2 dex on average) than those 
previously obtained by A10. 
Differences can be attributed to large differences in S/N between GTC 
and SDSS spectra. Moreover, it is worth noting that in the calculation 
of the nitrogen abundance we took into account the 
[\nii]$\lambda$6548\AA, which is clearly detected in the GTC spectra and 
mostly undetected in the SDSS spectra.
The three GP galaxies studied here show N/O ratios that 
are clearly higher than the usual upper limits of the {\it plateau} 
on the N/O vs. O/H diagram and more compatible with those of 
emission-line galaxies with higher metallicities and similar 
sSFR \citep[e.g.,][]{Mallery07} for which secondary nitrogen production 
takes over. 
Using a large sample of galaxies that include the GPs of C09, Iz11 
noticed that for galaxies with EW(H$\beta$)$<$100\AA\ there is a dependence 
of the N/O ratio on the total galaxy mass, implying some secondary nitrogen 
production in the highest-mass galaxies ($\ga$10$^9$M$_{\odot}$, see their 
Fig.~12). 

In order to explain the observed high N/O ratios in metal-poor SFGs 
several mechanisms, including time delays on chemical enrichment, effects 
of gas flows, and variations in the SFHs 
(e.g., recurrent bursts) have been proposed in the literature 
\citep[see e.g.,][]{Garnett90,Pilyugin92,Koppen99,Vanzee06,Molla06,Pilyugin12}.
Using a large sample of metal-poor emission-line galaxies, 
\citet{Izotov06} found an apparent increase of N/O with decreasing EW(\hb), 
best seen at intermediate metallicities. They interpreted this trend 
as an evidence of gradual enrichment of the H{\sc ii} regions in nitrogen 
by massive stars from the most recent starburst. 
In this line, the discovery of WR stars in the three GPs studied here 
offer additional ingredients to the discussion. 

Localized chemical pollution by WR stars has been argued to explain the 
enhancement in nitrogen abundance in the central starburst of nearby WR 
galaxies \citep[e.g., NGC5253][]
{Kobulnicky97,LopezSanchez07,LopezSanchez11,James09,Monreal10}.
These regions, with sizes of about tenths of parsecs usually show N/O 
enhancements consistent with few WRs stars \citep{Monreal10}. 
In their integrated spectra, most WR galaxies with EW(\hb) $\la$ 100\AA\ 
in \citet{Brinchmann08} \citep[see also][]{LopezSanchez10} show higher 
N/O than non-WR galaxies, suggesting a rapid enrichment by WR stars. 
However, most of these WR galaxies with N/O values comparable to our 
GPs show much higher oxygen, neon, and argon abundances at the same EW(\hb) 
(see their Fig.~15). 
Helium abundances in the three GP galaxies are roughly the same and {not} 
especially high. Indeed, these values are comparable to those found in 
most BCDs \citep[e.g., ][]{Pagel92}. 
As pointed out by \citet{Pagel86,Pagel92} (see also \citet{Kobulnicky97}) 
high He/H ratios are to be expected in regions with strong WR pollution. 
However, as remarked by \citet{Brinchmann08}, following results found in 
Galactic ring nebulae by \citet{Esteban92}, the expected average increase 
in He/H could be up to 0.7 dex lower than the expected increase in N/O. 
So the apparently normal He/H seen in the GPs may not give additional clues 
in this respect.

Finally, recent results from integral field spectroscopy on three 
(HS~0128+2832,HS~0837+4717 and Mrk930) nitrogen overabundant BCDs by 
P\'erez-Montero et al. (2011), have shown that these 
galaxies display high N/O, which are constant at spatial scales of the 
order of several kpc. Interestingly, their N/O values are rather 
similar to those of our target GPs. 
P\'erez-Montero et al. show the evolution of N/O with time for spherical 
gaseous distributions of constant density with initial 
12$+$log(O/H)$=$8 and log(N/O)$=$-1.6 , which is polluted 
by the stellar winds coming from stellar clusters with $Z=0.004$ and 
different stellar masses (from 10$^{7}$ to 10$^{10} M_{\odot}$). 
Importantly, their results show clearly that the number of WR stars 
estimated from spectroscopy cannot be responsible for the observed 
enhancement in the N abundance across the observed spatial scales. 
According to their results, in the case of the GPs it would be 
neccesary ionizing clusters with masses 2 or 3 orders of 
magnitude higher to reach 
such degree of pollution in the spatial scale 
covered by our spectral aperture (see \S3.3.1). 
This suggests that another chemical evolution scenario is required 
to explain the observed properties of the GPs.

One possiblity is to think our galaxies as (chemically) evolved enough 
to have reach the secondary nitrogen regime. 
\citet{Koppen05} used models to investigate the chemical evolution 
of galaxies during an episode of massive and sudden  
accretion of metal-poor gas due to, e.g., interaction with gas clouds 
or small gas-rich companions. 
In galaxies that has reached the secondary nitrogen regime a rapid 
decrease of the oxygen abundance can be produced during the infall, 
remaining the N/O unchanged, before a slower evolution which leads 
back to the closed-box relation. 
This theoretical model and similar results from numerical 
simulations \citep[e.g.,][]{Finlator08} led to A10 to suggest a 
scenario where the GPs would be experiencing the immediate  
effects of a recent, massive accretion of metal-poor gas 
(from the galaxy outskirt or beyond), possibly modulated by enriched 
outflows. 
This, at the same time, appears consistent with the SFH 
and observed morphology of the GP galaxies.  

\section{Summary and final remarks}
\label{s5}

In this paper, we have presented for the first time deep optical
broad-band imaging and spectroscopy for a sample of three GPs 
with a 10m-class telescope, the GTC.

Chemical abundances and physical properties of the ionized gas were 
derived with high precision in the three galaxies, confirming
previous findings. In particular, our results support several of the
conclusions drawn by A10: We find that these galaxies are metal poor
(7.8$\la$12+log(O/H)$\la$8.0) just like nearby BCDs. However, in
contrast to most BCDs, the studied GPs show a remarkably high N/O ratio
for their low oxygen abundance.  This is in agreement with previous
work by A10 \citep[see also][]{Pilyugin12} who claimed after a 
re-analysis of SDSS data for the C09 sample of GPs that a significant 
fraction of these systems shows enhanced nitrogen abundances. 

The high S/N GTC spectra allowed to discover clear signatures of WR 
stars in the three galaxies. One of the implications of this finding  
is that these WR stars might produce a localized chemical pollution, 
and therefore explain the apparent nitrogen over-abundance of the 
galaxies. However, by comparing with similar results for BCDs 
presented by \citet{PM11}, the derived ratio between WR and O stars 
for GPs seem to be not enough to explain the apparent nitrogen 
over-abundance as due to local pollution by WRs.   

With the aim of gaining further insight into the SFH of GPs, 
we carried out a pilot spectral synthesis study of these systems 
using the code {\sc starlight} in conjunction with the new-generation 
SSP library {\sc popstar} which includes nebular emission ({\sc run1}).  
As a consistency check, we additionally employed {\sc starlight} with 
purely stellar SSPs ({\sc run2}), as well as an evolutionary 
synthesis code that self-consistently accounts for Balmer emission line 
equivalent widths (EWs) for a simplified SFH comprising an old and a 
young stellar population ({\sc run3}).  The central outcome from the 
{\sc starlight}+{\sc popstar} model, qualitatively confirmed through the 
{\sc run2}\&{\sc run3}, is that the GPs under study currently undergo a
major starbust producing between $\sim$4\% and 20\% of their stellar
mass. This result is supported by the detection, for the first time in
these galaxies, of clear signatures of a significant number of WR
stars in their spectra.  The derived SFH of these systems shows large
discontinuities between a few $10^8$ yr and several Gyr, strongly
suggesting that their formation history is dominated by starbursts.
However, as models imply, GPs are old galaxies having had formed most
of their stellar mass several Gyr ago. The presence of an old stellar
component could be directly spectroscopically verified in one of our
GPs, adding further support to this conclusion.  Stellar masses
derived from {\sc run1} range between $1.7\times 10^9$ $M_{\odot}$ and 
$2.7\times 10^9$ $M_{\odot}$, they are thus fairly comparable to those
of luminous BCD galaxies in the nearby Universe.  Moreover, a surface
photometry study based on HST data indicates that our sample GPs
possess an extended low-surface brightness (LSB) envelope that, if due
to stellar emission, has structural properties compatible to those of 
luminous BCDs in the nearby Universe.

Clearly, to further elucidate the evolutionary scenarios of GPs we
need a better understanding of their chemical enrichment and star 
formation histories.  
In particular, it would be important to compare accurate
measurements of their chemical abundances, especially nitrogen, for a
larger sample of GPs with chemical evolution models taking into
account gas inflows and outflows.  High-quality spectra with 8-10m
class telescopes have been proved to be indispensable to study in
detail faint spectral features and to reduce uncertainties in the
analysis and interpretation of the data.

\acknowledgments

We would like to thank the anonymous referee for valuable 
comments and suggestions. 
We are very grateful to the GTC staff and, in particular, to 
A. Cabrera-Llavers for their unvaluable support with the OSIRIS 
observations and data reduction. 
This work has been funded by grants AYA2007-67965-C03-02, 
AYA2010-21887-C04-01, and 
CSD2006-00070: First Science with the GTC (\url{
  http://www.iac.es/consolider-ingenio-gtc/}) of the
Consolider-Ingenio 2010 Program, by the Spanish MICINN.
Polychronis Papaderos is supported by a Ciencia 2008 contract, funded 
by FCT/MCTES (Portugal) and POPH/FSE (EC).

Based on observations made with the Gran Telescopio Canarias (GTC),
instaled in the Spanish Observatorio del Roque de los Muchachos of
theInstituto de Astrofísica de Canarias, in the island of La Palma.




{\it Facilities:}\facility{GTC (OSIRIS)}, \facility{HST (WFPC2)}.

\begin{deluxetable}{lccccccccccccc}
\tabletypesize{\scriptsize}
\tablecaption{Main properties of the galaxies from the literature \label{tbl-1}}
\tablewidth{0pt}
\tablehead{
\colhead{ID} & \colhead{RA $J2000$} & \colhead{DEC $J2000$} & \colhead{z} &
\colhead{$m_g$}&\colhead{$M_g$}&\colhead{$E(B-V)$}&
\colhead{$r_{e}$}&\colhead{$L_{FUV}$}& 
\colhead{SFR$_{H\alpha+24}$} & 
\colhead{$M_{\star}$}&\colhead{12$+$log(O/H)}\\
\colhead{SDSS} &\colhead{deg}&\colhead{deg}&\colhead{}&\colhead{}& 
\colhead{} &\colhead{}&\colhead{kpc}&\colhead{$L_{\odot}$}&
\colhead{$M_{\odot}$ yr$^{-1}$}& \colhead{$M_{\odot}$}& \colhead{dex}
}
\startdata
J004054&10.22635&15.56938&0.283&20.86&-19.68&0.065&0.95&10.43&13.6&9.2\tablenotemark{a}&8.03\tablenotemark{a}     \\
J113303& 173.26578& 65.22815&0.241& 20.02& -20.16&0.011&0.77& 10.72& 7.7&9.30\tablenotemark{b}&7.97\tablenotemark{b},8.00\tablenotemark{c}       \\
J232539& 351.41345& 0.75200&0.277& 20.17& -20.32&0.037&0.81& 10.52& 12.8&9.41\tablenotemark{b}&8.29\tablenotemark{b},8.18\tablenotemark{c}    \\
\enddata
\tablecomments{Redshifts (z) were taken from SDSS DR7. $E(B-V)$ values are from \citet{Schlegel}. Magnitudes are not corrected from galactic and internal extinction. Absolute magnitudes were calculated as $M_g = m_g +$ 5 $-$ 5 $log(D_{L}) +$ 2.5 $log(1+z)$, where $D_{L}$ is the luminosity distance and $m_g$ is the Sloan $g'-$band apparent magnitude ($m_{petro}$). Half-light radius ($r_{e}$), obtained from WFPC2/HST optical broad-band imaging, and dust-corrected total SFRs are from \citet{Overzier09}. Logarithm of the FUV luminosity ($L_{FUV}$) was taken from \citet{Hoopes07}}
\tablenotetext{a}{From \citet{Overzier09}}
\tablenotetext{b}{From \citet{Izotov11}}
\tablenotetext{c}{Values obtained and used by \citet{Amorin10}. they were derived using the direct method in the case of GP113303, and the N2 calibration for GP232539. }
\end{deluxetable}

\begin{deluxetable}{cccccccl}
\tabletypesize{\scriptsize}
\tablecaption{OSIRIS-GTC log of observations \label{tbl-2}}
\tablewidth{0pt}
\tablehead{
\colhead{ID} & \colhead{CCD Chip} & \colhead{Slit\tablenotemark{a}/Filter} & 
\colhead{T$_{exp}$} & \colhead{Airmass} &
\colhead{Seeing} & \colhead{Date} & \colhead{Comment}
}
\startdata
GP004054&2&Sloan $z'$&2250s&1.10-1.25&$<$1.0&2010-07-08&clear/dark \\
\nodata &2&R1000B&3870s&1.1-1.2&1.2&2010-07-11&spectroscopic-dust/dark \\
\nodata &2&R1000R&3870s&1.13-1.26&1.0&2010-07-17&clear/dark \\
\nodata &2&R1000R&3870s&1.12-1.25&1.0-1.1&2010-08-02&clear-dust/dark  \\
GP113303&2&Sloan $z'$&2250s&1.45-1.60&1.0&2010-05-21&photometric/gray \\
\nodata &1&R1000B&3870s&1.35-1.45&$<$1.2&2010-06-03&photometric/gray \\
\nodata &1&R1000R&3870s&1.80-2.00&1.1-1.4&2010-07-09&clouds-dust/dark \\
GP232539&2& Sloan $z'$&2250s&1.3-1.5&$<$1.2&2010-07-07&clear-dust/dark \\
\nodata &1&R1000B&2580s&1.14&1.0&2010-07-09&cloudy-dust/dark \\
\nodata &2&R1000B&1290s&1.43&1.4&2010-07-12&cloudy-dust/dark \\
\nodata &1&R1000R&3870s&1.13-1.18&1.4&2010-07-10&spectroscopic-dust/dark \\
\enddata
\tablenotetext{a}{In all cases the spectroscopic observations were done with a 0.8 arcsec wide slit located in the parallactic angle.}
\end{deluxetable}


\begin{deluxetable}{lccccccc}
\tabletypesize{\scriptsize}
\tablecaption{Measured and extinction-corrected line fluxes and line identifications\label{tbl-3}}
\tablewidth{0pt}
\tablehead{
\colhead{$\lambda$, Line id.} & \colhead{f($\lambda$)}  & \multicolumn{2}{c}{GP004054} & \multicolumn{2}{c}{GP113303} & \multicolumn{2}{c}{GP232539} \\
 &  & \colhead{F($\lambda$)}  & \colhead{I($\lambda$)}  &  \colhead{F($\lambda$)} & \colhead{I($\lambda$)} &  \colhead{F($\lambda$)}  & \colhead{I($\lambda$)} 
}
\startdata
3188	HeI	&	0.47	&	20	$\pm$	2	&	28	$\pm$	4	&	--	 		&	--	 		&	32	$\pm$	14	&	51	$\pm$	22		\\
3499	HeI	&	0.37	&	11	$\pm$	2	&	14	$\pm$	2	&	--	 		&	--	 		&	--	 		&	--	 	 		\\
3531	HeI	&	0.36	&		 		&		$\pm$		&	25	$\pm$	2	&	30	$\pm$	5	&	--	 		&	--	 	 		\\
3697	H17	&	0.33	&	9	$\pm$	1	&	12	$\pm$	1	&	--	 		&	--	 		&	--	 		&	--	 	 		\\
3704	H16+HeI	&	0.33	&	11	$\pm$	2	&	14	$\pm$	3	&	15	$\pm$	3	&	17	$\pm$	4	&	8	$\pm$	1	&	11	$\pm$	2		\\
3712	H15	&	0.33	&	12	$\pm$	1	&	15	$\pm$	1	&	--	 		&	--	 		&	10	$\pm$	5	&	13	$\pm$	7		\\
3727	[OII]	&	0.32	&	1314	$\pm$	4	&	1680	$\pm$	87	&	1443	$\pm$	14	&	1675	$\pm$	212	&	1815	$\pm$	8	&	2479	$\pm$	111		\\
3750	H12	&	0.32	&	23	$\pm$	2	&	30	$\pm$	3	&	--	 		&	--	 		&	9	$\pm$	2	&	12	$\pm$	3		\\
3770	H11	&	0.31	&	31	$\pm$	4	&	39	$\pm$	5	&	21	$\pm$	5	&	24	$\pm$	6	&	14	$\pm$	3	&	19	$\pm$	4		\\
3798	H10	&	0.31	&	34	$\pm$	1	&	43	$\pm$	3	&	29	$\pm$	1	&	33	$\pm$	4	&	32	$\pm$	6	&	43	$\pm$	8		\\
3820	HeI	&	0.30	&	18	$\pm$	2	&	23	$\pm$	3	&	--	 		&	--	 		&	8	$\pm$	1	&	11	$\pm$	2		\\
3835	H9	&	0.30	&	56	$\pm$	2	&	71	$\pm$	4	&	52	$\pm$	3	&	60	$\pm$	8	&	29	$\pm$	3	&	39	$\pm$	5		\\
3868	[NeIII]	&	0.29	&	389	$\pm$	10	&	486	$\pm$	26	&	470	$\pm$	5	&	538	$\pm$	62	&	308	$\pm$	12	&	409	$\pm$	23		\\
3889	HeI+H8	&	0.29	&	150	$\pm$	4	&	186	$\pm$	10	&	154	$\pm$	4	&	176	$\pm$	20	&	143	$\pm$	5	&	189	$\pm$	10		\\
3916	NII 	&	0.28	&		 		&		$\pm$		&	68	$\pm$	11	&	78	$\pm$	15	&	--	 		&	--	 	 		\\
3968	[NeIII]+H7	&	0.27	&	287	$\pm$	4	&	351	$\pm$	16	&	252	$\pm$	6	&	285	$\pm$	31	&	182	$\pm$	6	&	235	$\pm$	12		\\
4026	[NII]+HeI	&	0.25	&	23	$\pm$	4	&	28	$\pm$	5	&	--	 		&	--	 		&	10	$\pm$	2	&	13	$\pm$	2		\\
4102	H$\delta$	&	0.23	&	247	$\pm$	3	&	294	$\pm$	11	&	263	$\pm$	10	&	293	$\pm$	29	&	218	$\pm$	6	&	272	$\pm$	11		\\
4243	[FeII]	&	0.19	&	--	 		&		$\pm$		&	25	$\pm$	3	&	27	$\pm$	4	&	--	 	--	&	--	 	 		\\
4340	H$\gamma$	&	0.16	&	463	$\pm$	4	&	522	$\pm$	14	&	418	$\pm$	4	&	449	$\pm$	28	&	437	$\pm$	5	&	508	$\pm$	13		\\
4363	[OIII]	&	0.15	&	79	$\pm$	3	&	88	$\pm$	4	&	84	$\pm$	3	&	90	$\pm$	6	&	64	$\pm$	11	&	74	$\pm$	12		\\
4471	HeI	&	0.12	&	40	$\pm$	1	&	44	$\pm$	1	&	37	$\pm$	6	&	39	$\pm$	7	&	33	$\pm$	3	&	37	$\pm$	4		\\
4571	MgI]	&	0.08	&	7	$\pm$	1	&	7	$\pm$	1	&	--	 	 	&	--	 	 	&	--	 	--	&	--	 	 		\\
4658	[FeIII]	&	0.06	&	18	$\pm$	2	&	19	$\pm$	2	&	18	$\pm$	4	&	18	$\pm$	4	&	18	$\pm$	2	&	19	$\pm$	2		\\
4686	HeII	&	0.05	&	14	$\pm$	2	&	14	$\pm$	2	&	--	 		&	--	 		&	--	 		&	--	 	 		\\
4702	[FeIII]	&	0.05	&	3	$\pm$	0	&	3	$\pm$	1	&	--	 		&	--	$\pm$		&	--	$\pm$		&	--	 	 		\\
4713	[ArIV]+HeI	&	0.04	&	12	$\pm$	1	&	13	$\pm$	1	&	9	$\pm$	2	&	9	$\pm$	2	&	5	$\pm$	1	&	5	$\pm$	1		\\
4740	[ArIV] 	&	0.03	&	9	$\pm$	1	&	9	$\pm$	1	&	--	 	 	&	--	 	 	&	--	 	 	&	--	 	 		\\
4815	[FeII]	&	0.01	&	6	$\pm$	1	&	6	$\pm$	1	&	16	$\pm$	5	&	16	$\pm$	5	&	--	 		&	--	 	 		\\
4861	H$\beta$	&	0.00	&	1000	$\pm$	3	&	1000	$\pm$	3	&	1000	$\pm$	6	&	1000	$\pm$	6	&	1000	$\pm$	5	&	1000	$\pm$	5		\\
4881	[FeIII]	&	0.00	&	--	 	 	&	--	$\pm$	--	&	19	$\pm$	2	&	19	$\pm$	2	&	7	$\pm$	1	&	7	$\pm$	1		\\
4921	HeI	&	-0.02	&	13	$\pm$	3	&	13	$\pm$	3	&	--	 	 	&	--	 	 	&	4	$\pm$	2	&	4	$\pm$	2		\\
4959	[OIII]	&	-0.03	&	2017	$\pm$	8	&	1978	$\pm$	11	&	1918	$\pm$	11	&	1895	$\pm$	22	&	1404	$\pm$	8	&	1369	$\pm$	9		\\
4986	[FeIII]	&	-0.03	&	14	$\pm$	2	&	13	$\pm$	2	&	14	$\pm$	6	&	13	$\pm$	6	&	12	$\pm$	1	&	12	$\pm$	1		\\
5007	[OIII]	&	-0.04	&	6080	$\pm$	13	&	5907	$\pm$	38	&	5499	$\pm$	27	&	5404	$\pm$	84	&	4209	$\pm$	24	&	4058	$\pm$	31		\\
5199	[NI]	&	-0.08	&	12	$\pm$	2	&	11	$\pm$	1	&	--	 	 	&	--	 	 	&	30	$\pm$	3	&	28	$\pm$	3		\\
5876	HeI	&	-0.20	&	131	$\pm$	3	&	112	$\pm$	5	&	128	$\pm$	2	&	117	$\pm$	9	&	141	$\pm$	5	&	116	$\pm$	6		\\
6046	[OI]	&	-0.23	&	31	$\pm$	18	&	26	$\pm$	16	&	--	 	 	&	--	 	 	&	--	 	--	&	--	 	 		\\
6300	[OI]	&	-0.26	&	55	$\pm$	2	&	45	$\pm$	3	&	82	$\pm$	9	&	72	$\pm$	11	&	87	$\pm$	8	&	68	$\pm$	7		\\
6312	[SIII]	&	-0.26	&	26	$\pm$	6	&	21	$\pm$	5	&	43	$\pm$	6	&	38	$\pm$	7	&	26	$\pm$	8	&	20	$\pm$	6		\\
6364	[OI]	&	-0.27	&	15	$\pm$	1	&	12	$\pm$	1	&	--	 		&	--	 		&	28	$\pm$	7	&	22	$\pm$	5		\\
6548	[NII] 	&	-0.30	&	64	$\pm$	10	&	51	$\pm$	8	&	21	$\pm$	10	&	18	$\pm$	9	&	38	$\pm$	11	&	28	$\pm$	9		\\
6563	H$\alpha$	&	-0.30	&	3588	$\pm$	12	&	2861	$\pm$	138	&	2952	$\pm$	33	&	2573	$\pm$	301	&	3878	$\pm$	49	&	2907	$\pm$	125		\\
6584	[NII] 	&	-0.30	&	148	$\pm$	6	&	117	$\pm$	7	&	172	$\pm$	6	&	150	$\pm$	20	&	332	$\pm$	4	&	248	$\pm$	11		\\
6678	HeI	&	-0.31	&	41	$\pm$	1	&	32	$\pm$	2	&	141	$\pm$	10	&	122	$\pm$	17	&	41	$\pm$	2	&	30	$\pm$	2		\\
6717	[SII]	&	-0.32	&	193	$\pm$	4	&	152	$\pm$	8	&	157	$\pm$	10	&	136	$\pm$	19	&	330	$\pm$	5	&	243	$\pm$	11		\\
6731	[SII]	&	-0.32	&	149	$\pm$	3	&	117	$\pm$	6	&	165	$\pm$	21	&	142	$\pm$	25	&	253	$\pm$	8	&	185	$\pm$	10		\\
7065	HeI	&	-0.36	&	43	$\pm$	8	&	33	$\pm$	6	&	--	 	 	&	--	 	 	&	37	$\pm$	13	&	26	$\pm$	9		\\
7083	[ArI]	&	-0.37	&	8	$\pm$	1	&	6	$\pm$	1	&	--	 	 	&	--	 	 	&	28	$\pm$	2	&	20	$\pm$	2		\\
7136	[ArIII]	&	-0.37	&	91	$\pm$	6	&	69	$\pm$	6	&	--	 	 	&	--	 	 	&	84	$\pm$	4	&	59	$\pm$	4		\\
7388	[Fe II]	&	-0.41	&	--	 		&	--	 		&	--	 		&	--	 		&	17	$\pm$	1	&	11	$\pm$	1		\\
7499	HeI	&	-0.42	&	67	$\pm$	2	&	48	$\pm$	4	&	--	 		&	--	 		&	13	$\pm$	2	&	9	$\pm$	1		\\
\tableline
c(H$\beta$)  &     &  \multicolumn{2}{c}{0.33 $\pm$ 0.07} & \multicolumn{2}{c}{0.20 $\pm$ 0.17} & \multicolumn{2}{c}{0.42 $\pm$ 0.05} \\
-W(H$\beta$)  (\AA)   &  &   \multicolumn{2}{c}{156} &  \multicolumn{2}{c}{74} & \multicolumn{2}{c}{60}  \\ 
F(H$\beta$) &    &  \multicolumn{2}{c}{18.7}  & \multicolumn{2}{c}{15.4} & \multicolumn{2}{c}{20.2} \\
\tableline
\enddata
\tablecomments{Wavelengths in Col.~1 ($\lambda$) are rest-frame. Measured [F($\lambda$)] and extinction-corrected [I($\lambda$)] emission-line fluxes are relative to F(H$\beta$) =  I(H$\beta$) = 1000 with their corresponding errors. We also give for each galaxy the corresponding constants of reddening [c(H$\beta$)], the H$\beta$ emission-line extinction-corrected total flux in units of 10$^{-14}$ erg s$^{-1}$ cm$^2$ \AA$^{-1}$, and the H$\beta$ equivalent width.}
\end{deluxetable}

\begin{deluxetable}{lccc}
\tabletypesize{\footnotesize}
\tablecaption{Derived physical conditions and chemical abundances\label{tbl-4}}
\tablewidth{0pt}
\tablehead{
\colhead{Object}&\colhead{GP004054}&\colhead{GP113303}&\colhead{GP232539} 
}
\startdata
n$_e$([S{\sc ii}]) (cm$^{-3}$)   &   $<$ 260  &  $<$ 680  &  $<$ 210 \\
t$_e$([O{\sc iii}]) (10$^4$ K)  &  1.34 $\pm$ 0.02 & 1.39 $\pm$ 0.04 & 1.46 $\pm$ 0.11 \\
t$_e$([O{\sc ii}])\tablenotemark{a}(10$^4$ K) &  1.52 $\pm$ 0.02 & 1.60 $\pm$ 0.02 & 1.60 $\pm$ 0.07 \\
t$_e$([N{\sc ii}]) (10$^4$ K)  &  1.26 $\pm$ 0.01 & 1.29 $\pm$ 0.02 & 1.32 $\pm$ 0.04 \\
t$_e$([S{\sc iii}])\tablenotemark{b} (10$^4$ K) & 1.27 $\pm$ 0.15 & 1.34 $\pm$ 0.16 & 1.42 $\pm$ 0.20 \\
\tableline

y$^+$  (4471) &  0.090 $\pm$ 0.002 & 0.078 $\pm$ 0.014 & 0.077 $\pm$ 0.009 \\
y$^+$  (5876) &  0.088 $\pm$ 0.004 & 0.086 $\pm$ 0.007 & 0.090 $\pm$ 0.004 \\
y$^+$  (6678) &  0.088 $\pm$ 0.006 & -- & 0.084 $\pm$ 0.007 \\
y$^+$  (7065)\tablenotemark{c} &  0.110 $\pm$ 0.020 & -- & 0.082 $\pm$ 0.028 \\
y$^{2+}$  (4686)  & 0.0012 $\pm$ 0.0001 & -- & -- \\
{\bf y}  & 0.089 $\pm$ 0.002 & 0.089 $\pm$ 0.004 & 0.087 $\pm$ 0.004 \\
\tableline

12+log(O$^+$/H$^+$)  & 7.19 $\pm$ 0.04 & 7.15 $\pm$ 0.07 & 7.30 $\pm$ 0.10 \\
12+log(O$^{+2}$/H$^+$)  & 7.91 $\pm$ 0.02 & 7.82 $\pm$ 0.03 & 7.65 $\pm$ 0.05 \\
{\bf 12+log(O/H)}  & 7.98 $\pm$ 0.06 & 7.91 $\pm$ 0.10 & 7.81 $\pm$ 0.14 \\
\tableline

12+log(N$^+$/H$^+$)  & 6.13 $\pm$ 0.01 & 6.11 $\pm$ 0.02 & 6.31 $\pm$ 0.04 \\
ICF(N$^+$) & 6.18 $\pm$ 0.18 & 5.73 $\pm$ 0.23 & 3.21 $\pm$ 0.59 \\
{\bf 12+log(N/H)} & 6.92 $\pm$ 0.02  & 6.87 $\pm$ 0.03 & 6.82 $\pm$ 0.08 \\
{\bf log(N/O)} & -1.06 $\pm$ 0.04 & -1.04 $\pm$ 0.08 & -0.99 $\pm$ 0.11 \\
\tableline

12+log(Ne$^{+2}$/H$^+$)  & 7.27 $\pm$ 0.03 & 7.26 $\pm$ 0.06 & 7.08 $\pm$ 0.08 \\
ICF(Ne$^{+2}$) & 1.08 $\pm$ 0.04 & 1.08 $\pm$ 0.06 & 1.11 $\pm$ 0.11 \\
{\bf 12+log(Ne/H)} & 7.30 $\pm$ 0.03  & 7.30 $\pm$ 0.06 & 7.13 $\pm$ 0.09 \\
{\bf log(Ne/O)} & -0.68 $\pm$ 0.07 & -0.61 $\pm$ 0.12 & -0.68 $\pm$ 0.17 \\
\tableline

12+log(Ar$^{+2}$/H$^+$)  & 5.54 $\pm$ 0.04 & -- & 5.40 $\pm$ 0.04 \\
12+log(Ar$^{+3}$/H$^+$)  & 4.57 $\pm$ 0.05 & -- & --  \\
ICF(Ar) & 1.04 $\pm$ 0.04 & --  & 3.00 $\pm$ 1.44 \\
{\bf 12+log(Ar/H)} & 5.60 $\pm$ 0.10  & --  & 5.88 $\pm$ 0.22 \\
{\bf log(Ar/O)} & -2.31 $\pm$ 0.10 & -- & -1.77 $\pm$ 0.23 \\
\tableline

12+log(S$^+$/H$^+$)  & 4.74 $\pm$ 0.15 & 4.76 $\pm$ 0.25 & 4.94 $\pm$ 0.14 \\
12+log(S$^{+2}$/H$^+$)  & 6.32 $\pm$ 0.25 & 6.49 $\pm$ 0.24 & 6.13 $\pm$ 0.27  \\
ICF(S) & 1.51 $\pm$ 0.40 & 1.55 $\pm$ 0.55  & 1.27 $\pm$ 0.27 \\
{\bf 12+log(S/H)} & 6.51 $\pm$ 0.35  & 6.69 $\pm$ 0.62  & 6.26 $\pm$ 0.36 \\
{\bf log(S/O)} & -1.48 $\pm$ 0.36 & -1.22 $\pm$ 0.43 & -1.55 $\pm$ 0.39 \\
\tableline

12+log(Fe$^{2+}$ & 5.69 $\pm$ 0.05 & 5.62 $\pm$ 0.09 & 5.59 $\pm$ 0.08 \\
ICF(Fe$^{2+}$  &  6.3 $\pm$ 4.8 & 7.0 $\pm$ 6.0 & 3.7 $\pm$ 2.7 \\
{\bf 12+log(Fe/H)} & 6.49 $\pm$ 0.25 & 6.39 $\pm$ 0.36 & 6.11 $\pm$ 0.29 \\
{\bf log(Fe/O)} & -1.49 $\pm$ 0.26 & -1.52 $\pm$ 0.38 & -1.69 $\pm$ 0.32 \\

\tableline
\enddata
\tablenotetext{a}{Derived from t$_e$([O{\sc iii}]) using the 
models described in \citet{PMyD03} and \citet{PMC09} for t$_e$([O{\sc ii}]) 
and t$_e$([N{\sc ii}]), respectively} 
\tablenotetext{b}{Derived from t$_e$([O{\sc iii}]) using the empirical 
relation described in \citet{Hagele06}}
\tablenotetext{c}{He{\sc i} 7065\AA\ should be affected by radiative 
transfer effects. However, we considered here their measurements in our 
calculations since $y^+$(7065) values are, within the uncertanties, in 
agreement with the $y^+$ values derived from the other helium lines.}
\end{deluxetable}

\clearpage
\newpage

\begin{deluxetable}{lccccccc}
\tabletypesize{\scriptsize}
\tablecaption{Model results from {\sc starlight} {\sc (run 1 and run 2)} \label{tbl-5}}
\tablecolumns{7} 
\tablewidth{0pt}
\tablehead{
\colhead{Galaxy}&\multicolumn{3}{c}{nebular continuum\tablenotemark{a} (\%)}& 
\colhead{$M_{\star,\rm young}$}&\colhead{$M_{\star,\rm total}$}&\colhead{$\chi^{2}$} \\
\colhead{}&\colhead{[\oii]}&\colhead{\hb}&\colhead{\ha}&\colhead{(\%)}&
\colhead{(log $M_{\odot}$)}&\colhead{}
}
\startdata
\sidehead{\sc run 1: popstar models}
GP004054& 10& 18& 35& 22&9.18&1.05\\
GP113303& 3.5& 6& 11& 4.2&9.44&0.86\\
GP232539& 8& 10.5& 20& 11.3&9.38&1.54\\
\sidehead{\sc run 2: B\&C models}
GP004054& \nodata& \nodata& \nodata& 15.7& 9.22 & 1.17 \\
GP113303& \nodata& \nodata& \nodata& 1.9 & 9.68 & 0.95 \\
GP232539& \nodata& \nodata& \nodata& 5.4 & 9.90 & 1.37 \\

\enddata
\tablenotetext{a}{For three different wavelength regions 
(around [\oii], \hb, and \ha), columns 2, 3, and 4 give the 
estimated fraction nebular continuum emission derived from 
{\sc starlight} fits to each galaxy spectrum using {\sc popstar} 
models (see text)}
\end{deluxetable}

\begin{deluxetable}{lccccccc}
\tabletypesize{\scriptsize}
\tablecaption{Results from the two-component evolutionary synthesis
  model (\sc run 3) \label{tbl-6}}
\tablewidth{0pt}
\tablehead{
\colhead{Galaxy}&\colhead{$t_{\rm burst}$}&\colhead{$b_{par}$}&
\colhead{$C(\hb)$}& \colhead{EW(\ha)}&\colhead{EW(\hb)}&
\colhead{$M_{\star,\rm young}$} & \colhead{$\chi^{2}$}\\
\colhead{}&\colhead{(Myr)}&\colhead{}&\colhead{}&\colhead{(\AA)}&
\colhead{(\AA)}&\colhead{(\%)} & 
}
\startdata
GP004054& 4 & 5.3 & 0.41 & 894 & 139 & 8.3 &  1.5\\
GP113303& 5 & 4.2 & 0.32 & 445 & 76  & 6.7 & 2.1\\
GP232539& 5 & 2.8 & 0.43 & 344 & 53  & 4.4 &  1.4\\
\enddata
\tablecomments{Column 2: age of the young stellar component. 
Column 3: burst parameter. 
Column 4: reddening constant predicted by the evolutionary model. 
Columns 5 and 6: 
predicted equivalent widths of \ha\ and \hb\ emission lines.
}
\end{deluxetable}

\begin{deluxetable}{lccccc}
\tabletypesize{\scriptsize}
\tablecaption{Wolf-Rayet features detected in the galaxy sample \label{tbl-7}}
\tablewidth{0pt}
\tablehead{
\colhead{Galaxy}& \colhead{$I_{\rm BB}$($\lambda$)\tablenotemark{a}}& 
\colhead{log$L_{\rm BB}$\tablenotemark{b}}&\colhead{$EW_{\rm BB}$}&
\colhead{N(WR$_{\star}$)\tablenotemark{c}}&
\colhead{N(WR$_{\star}$/O$_{\star}$)\tablenotemark{d}} 
}
\startdata
GP004054&  34$\pm$4 &40.1&  4.2$\pm$0.5&  806$\pm$97 & 0.011$\pm$0.001  \\
GP113303&  80$\pm$15&40.8&  5.5$\pm$1.0& 1239$\pm$228& 0.029$\pm$0.005  \\
GP232539&  36$\pm$6 &40.3&  1.9$\pm$0.3& 1145$\pm$191& 0.012$\pm$0.002  \\
\enddata
\tablenotetext{a}{Extinction-corrected, emission-line flux measured for 
the WR blue bump relative to I($\hb$)$=$1000 with their corresponding 
errors.}
\tablenotetext{b}{Luminosity of the WR blue bump calculated using 
$I_{\rm BB}$($\lambda$) and the distances adopted throughout the paper}
\tablenotetext{c}{Number of WR stars as derived from $L_{\rm BB}$ and 
adopting a mean luminosity for a WR star log$L_{\rm WR\star}=37.248$ 
from {\sc starburst99} models with stellar metallicity $Z=0.004$ 
\citep{PM10} }
\tablenotetext{d}{WR to O star number ratio and its errors. 
The number of  O stars were derived from $L_{\rm H\alpha}$ and adopting 
a mean \ha\ luminosity for an O star log$L_{\rm H\alpha, O\star}=37.21$.}
\end{deluxetable}

\begin{thebibliography}{}

\bibitem[Adamo et al.(2011)]{Adamo11} Adamo, A., {\"O}stlin, 
G., Zackrisson, E., et al.\ 2011, \mnras, 415, 2388 

\bibitem[Allen et al.(1976)]{Allen76} Allen, D.~A., Wright, 
A.~E., \& Goss, W.~M.\ 1976, \mnras, 177, 91 

\bibitem[Alloin et al.(1979)]{Alloin79} Alloin, D., Collin-Souffrin, S., Joly, M., \& Vigroux, L.\ 1979, \aap, 78, 200 

\bibitem[Amor{\'{\i}}n et al.(2007)]{Amorin07} Amor{\'{\i}}n, R.~O., Mu{\~n}oz-Tu{\~n}{\'o}n, C., Aguerri, J.~A.~L., Cair{\'o}s, L.~M., \& Caon, N.\ 2007, \aap, 467, 541 

\bibitem[Amor{\'{\i}}n et al.(2009)]{Amorin09} Amor{\'{\i}}n, R., Aguerri, J.~A.~L., Mu{\~n}oz-Tu{\~n}{\'o}n, C., \& Cair{\'o}s, L.~M.\ 2009, \aap, 501, 75 

\bibitem[Amor{\'{\i}}n et al.(2010)]{Amorin10} Amor{\'{\i}}n, R.~O., P{\'e}rez-Montero, E., \& V{\'{\i}}lchez, J.~M.\ 2010, \apjl, 715, L128 

\bibitem[Amor\'in et al.(2011)]{Amorin11} Amor\'in, R.~O., V\'ilchez, J.~M., \& P\'erez-Montero, E., 
2011, in JENAM Symposium {\sl Dwarf Galaxies: Keys to Galaxy
Formation and Evolution}, P. Papaderos, S.  Recchi \& G. Hensler (eds.).
Lisbon, September 2010, Springer Verlag, in press

\bibitem[Asplund et al.(2009)]{Asplund09} Asplund, M., Grevesse, N., 
Sauval, A.~J., \& Scott, P.\ 2009, ARA\&A, 47, 481 

\bibitem[Basu-Zych et al.(2007)]{Basu07} Basu-Zych, A.~R., et 
al.\ 2007, \apjs, 173, 457 

\bibitem[Bauer et al.(2005)]{Bauer05} Bauer, A.~E., Drory, N., 
Hill, G.~J., \& Feulner, G.\ 2005, \apjl, 621, L89 

\bibitem[Berg et al.(2011)]{Berg11} Berg, D.~A., Skillman, 
E.~D., \& Marble, A.~R.\ 2011, \apj, 738, 2 

\bibitem[Bergvall \& \"Ostlin(2002)]{BergvallOstlin02}Bergvall, N. \& \"Ostlin, G. 2002, A\&A, 390, 891

\bibitem[Bournaud \& Elmegreen (2009)]{Bournaud09} Bournaud, F., \& Elmegreen, B.~G.\ 2009, \apj, 694, L158

\bibitem[Brinchmann et al.(2004)]{Brinchmann04} Brinchmann, J., 
Charlot, S., White, S.~D.~M., Tremonti, C., Kauffmann, G., Heckman, T., 
\& Brinkmann, J.\ 2004, \mnras, 351, 1151 

\bibitem[Brinchmann et al.(2008)]{Brinchmann08} Brinchmann, J., Kunth, D., \& Durret, F.\ 2008, \aap, 485, 657 

\bibitem[Bruzual \& Charlot(2003)]{B&C03} Bruzual, G., \& Charlot, S.\ 2003, \mnras, 344, 1000 

\bibitem[Bundy et al.(2006)]{Bundy06} Bundy, K., et al.\ 2006, 
\apj, 651, 120

\bibitem[Campbell et al.(1986)]{Campbell86} Campbell, A., 
Terlevich, R., \& Melnick, J.\ 1986, \mnras, 223, 811 

\bibitem[Cair{\'o}s et al.(2001)]{Cairos01} Cair{\'o}s, L.~M., 
V{\'{\i}}lchez, J.~M., Gonz{\'a}lez P{\'e}rez, J.~N., Iglesias-P{\'a}ramo, 
J., \& Caon, N.\ 2001, \apjs, 133, 321 

\bibitem[Cair{\'o}s et al.(2003)]{Cairos03} Cair{\'o}s, L.~M., 
Caon, N., Papaderos, P., et al.\ 2003, \apj, 593, 312 

\bibitem[Cardamone et al.(2009)]{Cardamone09} Cardamone, C., et 
al.\ 2009, \mnras, 399, 1191 

\bibitem[Cardelli, Clayton, \& Mathis(1989)]{Cardelli89} Cardelli, 
J.~A., Clayton, G.~C., \& Mathis, J.~S.\ 1989, Apj, 345, 245 

\bibitem[Cepa et al.(2000)]{Cepa2000} Cepa, J., et al.\ 2000, 
\procspie, 4008, 623 

\bibitem[Cid Fernandes et al.(2004)]{Cid04} Cid Fernandes, 
R., Gu, Q., Melnick, J., Terlevich, E., Terlevich, R., Kunth, D., Rodrigues 
Lacerda, R., \& Joguet, B.\ 2004, \mnras, 355, 273 

\bibitem[Cid Fernandes et al.(2005)]{Cid05} Cid Fernandes, 
R., Mateus, A., Sodr{\'e}, L., Stasi{\'n}ska, G., 
\& Gomes, J.~M.\ 2005, \mnras, 358, 363 

\bibitem[Conti(1991)]{Conti91} Conti, P.~S.\ 1991, \apj, 377, 115 

\bibitem[Corbin et al.(2006)]{Corbin06} Corbin, M.~R., Vacca, 
W.~D., Cid Fernandes, R., Hibbard, J.~E., Somerville, R.~S., 
\& Windhorst, R.~A.\ 2006, \apj, 651, 861 

\bibitem[Cowie et al.(1996)]{Cowie96} Cowie, L.~L., Songaila, 
A., Hu, E.~M., \& Cohen, J.~G.\ 1996, \aj, 112, 839 

\bibitem[Cowie \& Barger(2008)]{Cowie08} Cowie, L.~L., \& Barger, A.~J.\ 2008, \apj, 686, 72 

\bibitem[Cresci et al. (2010)]{Cresci10} Cresci, G., Mannucci, F., Maiolino, R., Marconi, A., Gnerucci, A., \& Magrini, L.\ 2010, Natur, 467, 811

\bibitem[Crowther(2007)]{Crowther07} Crowther, P.~A.\ 2007, \araa, 45, 177 

\bibitem[Daddi et al. (2007)]{Daddi07} Daddi, E., et al.\ 2007, \apj, 670, 156  

\bibitem[D\'\i az(1988)]{Diaz88} D\'\i az, A.~I.\ 1988, MNRAS, 231, 57 

\bibitem[Erb et al.(2006)]{Erb06} Erb, D.~K., Shapley, A.~E., 
Pettini, M., Steidel, C.~C., Reddy, N.~A., 
\& Adelberger, K.~L.\ 2006, \apj, 644, 813 

\bibitem[Esteban et al.(1992)]{Esteban92} Esteban, C., Vilchez, J.~M., Smith, L.~J., \& Clegg, R.~E.~S.\ 1992, \aap, 259, 629 

\bibitem[Finkelstein et al.(2011)]{Finkelstein11} Finkelstein, S.~L., 
Hill, G.~J., Gebhardt, K., et al.\ 2011, \apj, 729, 140 

\bibitem[Finlator \& Dav{\'e}(2008)]{Finlator08} Finlator, K., \& Dav{\'e}, R.\ 2008, \mnras, 385, 2181 

\bibitem[Fioc \& Rocca-Volmerange(1997)]{Fioc97} Fioc, M., \& Rocca-Volmerange, B.\ 1997, \aap, 326, 950 

\bibitem[Gavil{\'a}n et al.(2006)]{Gavila06} Gavil{\'a}n, M., Moll{\'a}, M., \& Buell, J.~F.\ 2006, \aap, 450, 509 

\bibitem[Garcia-Vargas et al.(1997)]{GarciaVargas97} Garcia-Vargas, 
M.~L., Gonzalez-Delgado, R.~M., Perez, E., Alloin, D., Diaz, A., 
\& Terlevich, E.\ 1997, \apj, 478, 112 

\bibitem[Garnett(1990)]{Garnett90} Garnett, D.~R.\ 1990, \apj, 
363, 142 

\bibitem[Gil de Paz \& Madore(2005)]{GdP05} Gil de Paz, A., \& Madore, B.~F.\ 2005, \apjs, 156, 345 

\bibitem[Gilbank et al.(2011)]{Gilbank11} Gilbank, D.~G., et al.\ 
2011, \mnras, 402 

\bibitem[Gon{\c c}alves et al.(2010)]{Goncalves10} Gon{\c c}alves, 
T.~S., et al.\ 2010, \apj, 724, 1373 

\bibitem[Gonzalez-Delgado et al.(1994)]{GonzalezDelgado94} 
Gonzalez-Delgado, R.~M., et al.\ 1994, \apj, 437, 239 

\bibitem[Guseva et al.(2000)]{Guseva00} Guseva, N.~G., Izotov, 
Y.~I., \& Thuan, T.~X.\ 2000, \apj, 531, 776 

\bibitem[Guseva et al.(2001)]{Guseva01} Guseva, N.~G., Izotov, Y.~I., Papaderos, P., et al.\ 2001, \aap, 378, 756 

\bibitem[Guseva et al.(2007)]{Guseva07} Guseva, N.~G., Izotov, Y.~I., Papaderos, P., \& Fricke, K.~J.\ 2007, \aap, 464, 885 

\bibitem[Guseva et al.(2009)]{Guseva09} Guseva, N.~G., Papaderos, P., Meyer, H.~T., Izotov, Y.~I., \& Fricke, K.~J.\ 2009, \aap, 505, 63 

\bibitem[Guzman et al.(1997)]{Guzman97} Guzman, R., Gallego, J., 
Koo, D.~C., Phillips, A.~C., Lowenthal, J.~D., Faber, S.~M., Illingworth, 
G.~D., \& Vogt, N.~P.\ 1997, \apj, 489, 559 

\bibitem[Guzm{\'a}n et al.(2003)]{Guzman03} Guzm{\'a}n, R., 
{\"O}stlin, G., Kunth, D., Bershady, M.~A., Koo, D.~C., 
\& Pahre, M.~A.\ 2003, \apjl, 586, L45 

\bibitem[H{\"a}gele et al.(2006)]{Hagele06} H{\"a}gele, G.~F., 
P{\'e}rez-Montero, E., D{\'{\i}}az, {\'A}.~I., Terlevich, E., \& Terlevich, 
R.\ 2006, MNRAS, 372, 293 

\bibitem[H{\"a}gele et al.(2008)]{Hagele08} H{\"a}gele, G.~F., 
D{\'{\i}}az, {\'A}.~I., Terlevich, E., Terlevich, R., P{\'e}rez-Montero, 
E., \& Cardaci, M.~V.\ 2008, \mnras, 383, 209 

\bibitem[Heckman et al.(2005)]{Heckman05} Heckman, T.~M., et al.\ 2005, \apjl, 619, L35 

\bibitem[Henry et al.(2006)]{Henry06} Henry, R.~B.~C., Nava, 
A., \& Prochaska, J.~X.\ 2006, \apj, 647, 984 

\bibitem[Holtzman et al.(1995)]{Holtzman95}Holtzman et al. 1995, PASP, 107, 1065

\bibitem[Hoopes et al.(2007)]{Hoopes07} Hoopes, C.~G., et al.\ 
2007, \apjs, 173, 441 

\bibitem[Hoyos et al.(2005)]{Hoyos05} Hoyos, C., Koo, D.~C., 
Phillips, A.~C., Willmer, C.~N.~A., 
\& Guhathakurta, P.\ 2005, \apjl, 635, L21 

\bibitem[Hu et al.(1998)]{Hu98} Hu, E.~M., Cowie, L.~L., 
\& McMahon, R.~G.\ 1998, \apjl, 502, L99 

\bibitem[Izotov et al.(1997)]{Izotov97}Izotov et al. 1997, \apj, 476, 698

\bibitem[Izotov \& Thuan(1999)]{Izotov99} Izotov, Y.~I., \& Thuan, T.~X.\ 1999, \apj, 511, 639 

\bibitem[Izotov et al.(2006)]{Izotov06} Izotov, Y.~I., Stasi{\'n}ska, G., Meynet, G., Guseva, N.~G., \& Thuan, T.~X.\ 2006, \aap, 448, 955 

\bibitem[Izotov et al.(2011)]{Izotov11} Izotov, Y.~I., Guseva, 
N.~G., \& Thuan, T.~X.\ 2011, \apj, 728, 161 

\bibitem[James et al.(2009)]{James09} James, B.~L., Tsamis, 
Y.~G., Barlow, M.~J., Westmoquette, M.~S., Walsh, J.~R., Cuisinier, F., 
\& Exter, K.~M.\ 2009, \mnras, 398, 2 

\bibitem[Kakazu et al.(2007)]{Kakazu07} Kakazu, Y., Cowie, 
L.~L., \& Hu, E.~M.\ 2007, \apj, 668, 853 

\bibitem[Kehrig et al.(2006)]{Kehrig06} Kehrig, C., V{\'{\i}}lchez, J.~M., Telles, E., Cuisinier, F., \& P{\'e}rez-Montero, E.\ 2006, \aap, 457, 477 

\bibitem[Kehrig et al.(2008)]{Kehrig08} Kehrig, C., V{\'{\i}}lchez, J.~M., S{\'a}nchez, S.~F., et al.\ 2008, \aap, 477, 813 

\bibitem[Kobulnicky \& Skillman(1996)]{Kobulnicky96} Kobulnicky, H.~A., \& Skillman, E.~D.\ 1996, \apj, 471, 211 

\bibitem[Kobulnicky \& Skillman(1997)]{Kobulnicky97} Kobulnicky, H.~A., \& Skillman, E.~D.\ 1997, \apj, 489, 636 

\bibitem[Koo et al.(1994)]{Koo94} Koo, D.~C., Bershady, 
M.~A., Wirth, G.~D., Stanford, S.~A., 
\& Majewski, S.~R.\ 1994, \apjl, 427, L9 

\bibitem[K{\"o}ppen \& Edmunds(1999)]{Koppen99} K{\"o}ppen, J., \& Edmunds, M.~G.\ 1999, \mnras, 306, 317 

\bibitem[K{\"o}ppen \& Hensler(2005)]{Koppen05} K{\"o}ppen, J., \& Hensler, G.\ 2005, \aap, 434, 531 

\bibitem[Krueger et al.(1995)]{Krueger95} Krueger, H., Fritze-v.~Alvensleben, U., \& Loose, H.-H.\ 1995, \aap, 303, 41 

\bibitem[Kunth \& Sargent(1981)]{Kunth81} Kunth, D., \& Sargent, W.~L.~W.\ 1981, \aap, 101, L5 

\bibitem[Lee et al.(2004)]{Lee04} Lee, J.~C., Salzer, J.~J., 
\& Melbourne, J.\ 2004, \apj, 616, 752 

\bibitem[Leitherer et al.(1999)]{Leitherer99} Leitherer, C., et 
al.\ 1999, \apjs, 123, 3 

\bibitem[Lilly et al.(1996)]{Lilly96} Lilly, S.~J., Le Fevre, 
O., Hammer, F., \& Crampton, D.\ 1996, \apjl, 460, L1 

\bibitem[Lintott et al.(2008)]{Lintott08} Lintott, C.~J., et al.\ 
2008, \mnras, 389, 1179 

\bibitem[Lintott et al.(2011)]{Lintott11} Lintott, C., et al.\ 
2011, \mnras, 410, 166 

\bibitem[Lodders(2003)]{Lodders03} Lodders, K.\ 2003, \apj, 591, 1220 

\bibitem[L{\'o}pez-S{\'a}nchez et al.(2007)]{LopezSanchez07} 
L{\'o}pez-S{\'a}nchez, {\'A}.~R., Esteban, C., Garc{\'{\i}}a-Rojas, J., 
Peimbert, M., \& Rodr{\'{\i}}guez, M.\ 2007, \apj, 656, 168 

\bibitem[L{\'o}pez-S{\'a}nchez \& Esteban(2010)]{LopezSanchez10} L{\'o}pez-S{\'a}nchez, {\'A}.~R., \& Esteban, C.\ 2010, \aap, 516, A104 

\bibitem[L{\'o}pez-S{\'a}nchez et al.(2011)]{LopezSanchez11} 
L{\'o}pez-S{\'a}nchez, {\'A}.~R., Mesa-Delgado, A., 
L{\'o}pez-Mart{\'{\i}}n, L., \& Esteban, C.\ 2011, \mnras, 411, 2076 

\bibitem[Lowenthal et al.(1997)]{Lowenthal97} Lowenthal, J.~D., et 
al.\ 1997, \apj, 481, 673 

\bibitem[Luridiana et al.(2003)]{Luridiana03} Luridiana, V., 
Peimbert, A., Peimbert, M., \& Cervi{\~n}o, M.\ 2003, \apj, 592, 846 

\bibitem[Mallery et al.(2007)]{Mallery07} Mallery, R.~P., et al.\ 
2007, \apjs, 173, 482 

\bibitem[Mart{\'{\i}}n-Manj{\'o}n et al.(2010)]{Manjon10} 
Mart{\'{\i}}n-Manj{\'o}n, M.~L., Garc{\'{\i}}a-Vargas, M.~L., Moll{\'a}, 
M., \& D{\'{\i}}az, A.~I.\ 2010, \mnras, 403, 2012 

\bibitem[Masegosa et al.(1991)]{Masegosa91} Masegosa, J., Moles, M., \& del Olmo, A.\ 1991, \aap, 244, 273 

\bibitem[Masegosa et al.(1994)]{Masegosa94} Masegosa, J., Moles, 
M., \& Campos-Aguilar, A.\ 1994, \apj, 420, 576 

\bibitem[Mateus et al.(2006)]{Mateus06} Mateus, A., Sodr{\'e}, 
L., Cid Fernandes, R., Stasi{\'n}ska, G., Schoenell, W., 
\& Gomes, J.~M.\ 2006, \mnras, 370, 721 

\bibitem[Moll{\'a} et al.(2006)]{Molla06} Moll{\'a}, M., 
V{\'{\i}}lchez, J.~M., Gavil{\'a}n, M., 
\& D{\'{\i}}az, A.~I.\ 2006, \mnras, 372, 1069 

\bibitem[Moll{\'a} et al.(2009)]{Molla09} Moll{\'a}, M., 
Garc{\'{\i}}a-Vargas, M.~L., \& Bressan, A.\ 2009, \mnras, 398, 451 

\bibitem[Monreal-Ibero et al.(2010)]{Monreal10} Monreal-Ibero, A., V{\'{\i}}lchez, J.~M., Walsh, J.~R., \& Mu{\~n}oz-Tu{\~n}{\'o}n, C.\ 2010, \aap, 517, A27 

\bibitem[Nava et al.(2006)]{Nava06} Nava, A., Casebeer, D., 
Henry, R.~B.~C., \& Jevremovic, D.\ 2006, \apj, 645, 1076 

\bibitem[Noeske et al.(2003)]{Noeske03} Noeske, K.~G., Papaderos, P., Cair{\'o}s, L.~M., \& Fricke, K.~J.\ 2003, \aap, 410, 481 

\bibitem[Noeske et al.(2006)]{Noeske06} Noeske, K.~G., Koo, 
D.~C., Phillips, A.~C., Willmer, C.~N.~A., Melbourne, J., Gil de Paz, A., 
\& Papaderos, P.\ 2006, \apjl, 640, L143 

\bibitem[Noeske et al.(2007)]{Noeske07} Noeske, K.~G., et al.\ 
2007, \apjl, 660, L43 

\bibitem[Olive \& Skillman(2004)]{OliveySkill} Olive, K.~A., \& 
Skillman, E.~D.\ 2004, ApJ, 617, 29 

\bibitem[{\"O}stlin et al.(2001)]{Ostlin01} {\"O}stlin, G., Amram, P., 
Bergvall, N., et al.\ 2001, \aap, 374, 800 

\bibitem[Overzier et al.(2008)]{Overzier08} Overzier, R.~A., et 
al.\ 2008, \apj, 677, 37 

\bibitem[Overzier et al.(2009)]{Overzier09} Overzier, R.~A., et 
al.\ 2009, \apj, 706, 203 

\bibitem[Overzier et al.(2010)]{Overzier10} Overzier, R.~A., 
Heckman, T.~M., Schiminovich, D., Basu-Zych, A., Gon{\c c}alves, T., 
Martin, D.~C., \& Rich, R.~M.\ 2010, \apj, 710, 979 

\bibitem[Overzier et al.(2011)]{Overzier11} Overzier, R.~A., et 
al.\ 2011, \apjl, 726, L7 

\bibitem[Pagel et al.(1986)]{Pagel86} Pagel, B.~E.~J., 
Terlevich, R.~J., \& Melnick, J.\ 1986, \pasp, 98, 1005 

\bibitem[Pagel et al.(1992)]{Pagel92} Pagel, B.~E.~J., 
Simonson, E.~A., Terlevich, R.~J., 
\& Edmunds, M.~G.\ 1992, \mnras, 255, 325 

\bibitem[Papaderos et al.(1996)]{Papaderos96} Papaderos, P., Loose, H.-H., Fricke, K.~J., \& Thuan, T.~X.\ 1996, \aap, 314, 59 

\bibitem[Papaderos et al.(1998)]{Papaderos98} Papaderos, P., Izotov, Y.~I., Fricke, K.~J., Thuan, T.~X., \& Guseva, N.~G.\ 1998, \aap, 338, 43 

\bibitem[Papaderos et al.(2002)]{Papaderos02} Papaderos, P., Izotov, Y.I.,
  Thuan, T.X., Noeske, K.G., Fricke, K.J., Guseva, N.G., Green, R.F. 2002,
  \aap, 393, 461

\bibitem[Papaderos et al.(2008)]{Papaderos08} Papaderos, P., Guseva, N.~G., Izotov, Y.~I., \& Fricke, K.~J.\ 2008, \aap, 491, 113 

\bibitem[Pilyugin(1992)]{Pilyugin92} Pilyugin, L.~S.\ 1992, \aap, 260, 58 

\bibitem[P\'erez-Montero \& D\'iaz(2003)]{PMyD03} P\'erez-Montero, 
E. \& D\'iaz, A.I. 2003, MNRAS, 346, 105.

\bibitem[P{\'e}rez-Montero et al.(2006)]{PM06} 
P{\'e}rez-Montero, E., D{\'{\i}}az, A.~I., V{\'{\i}}lchez, J.~M., 
\& Kehrig, C.\ 2006, A\&A, 449, 193 

\bibitem[P{\'e}rez-Montero et al.(2007)]{PM07} P{\'e}rez-Montero E., 
H{\"a}gele G.~F., Contini T., D{\'{\i}}az {\'A}.~I., 2007, MNRAS, 381, 125 

\bibitem[P{\'e}rez-Montero et al.(2009)]{PM09} P{\'e}rez-Montero, E., Contini, T., Lamareille, F., et al.\ 2009, \aap, 495, 73 

\bibitem[P{\'e}rez-Montero \& Contini(2009)]{PMC09} P{\'e}rez-Montero, 
E., \& Contini, T.\ 2009, MNRAS, 398, 949 

\bibitem[P{\'e}rez-Montero et al.(2010)]{PM10} 
P{\'e}rez-Montero, E., Garc{\'{\i}}a-Benito, R., H{\"a}gele, G.~F., 
\& D{\'{\i}}az, {\'A}.~I.\ 2010, \mnras, 404, 2037 

\bibitem[P{\'e}rez-Montero et al.(2011)]{PM11} P{\'e}rez-Montero, E., V{\'{\i}}lchez, J.~M., Cedr{\'e}s, B., et al.\ 2011, \aap, 532, A141 

\bibitem[Pettini et al.(2001)]{Pettini01} Pettini, M., Shapley, 
A.~E., Steidel, C.~C., Cuby, J.-G., Dickinson, M., Moorwood, A.~F.~M., 
Adelberger, K.~L., \& Giavalisco, M.\ 2001, \apj, 554, 981 

\bibitem[Phillips et al.(1997)]{Phillips97} Phillips, A.~C., 
Guzman, R., Gallego, J., Koo, D.~C., Lowenthal, J.~D., Vogt, N.~P., Faber, 
S.~M., \& Illingworth, G.~D.\ 1997, \apj, 489, 543 

\bibitem[Pilyugin(1993)]{Pilyugin93} Pilyugin, L.~S.\ 1993, \aap, 277, 42 

\bibitem[Pilyugin et al.(2003)]{Pilyugin03} Pilyugin, L.~S., Thuan, T.~X., \& V{\'{\i}}lchez, J.~M.\ 2003, \aap, 397, 487 

\bibitem[Pilyugin et al.(2012)]{Pilyugin12} Pilyugin, L.~S., 
Vilchez, J.~M., Mattsson, L., \& Thuan, T.~X.\ 2012, arXiv:1201.1554 

\bibitem[Primack et al.(1998)]{Primack98} Primack, J.~R., 
Somerville, R.~S., Faber, S.~M., 
\& Wechsler, R.~H.\ 1998, \physrep, 307, 15 

\bibitem[Renzini \& Voli(1981)]{Renzini81} Renzini, A., \& Voli, M.\ 1981, \aap, 94, 175 

\bibitem[Reverte et al.(2007)]{Reverte07} Reverte, D., 
V{\'{\i}}lchez, J.~M., Hern{\'a}ndez-Fern{\'a}ndez, J.~D., 
\& Iglesias-P{\'a}ramo, J.\ 2007, \aj, 133, 705 

\bibitem[Rodr{\'{\i}}guez \& Rubin(2004)]{RyR04} Rodr{\'{\i}}guez, M., 
\& Rubin, R.~H.\ 2004, Recycling Intergalactic and Interstellar Matter, 
217, 188 

\bibitem[Salim et al.(2007)]{Salim07} Salim, S., et al.\ 2007, 
\apjs, 173, 267 

\bibitem[Salzer et al.(2009)]{Salzer09} Salzer, J.~J., Williams, 
A.~L., \& Gronwall, C.\ 2009, \apjl, 695, L67 

\bibitem[Sargent \& Searle(1970)]{S&S70} Sargent, W.~L.~W., \& Searle, L.\ 1970, \apjl, 162, L155 

\bibitem[Schaerer \& Vacca(1998)]{SchaererVacca98} Schaerer, D., \& Vacca, W.~D.\ 1998, \apj, 497, 618 

\bibitem[Schaerer et al.(1999)]{Schaerer99} Schaerer, D., Contini, T., 
\& Pindao, M.\ 1999, \aaps, 136, 35 

\bibitem[Schlegel et al.(1998)]{Schlegel} Schlegel, D.~J., 
Finkbeiner, D.~P., \& Davis, M.\ 1998, \apj, 500, 525 

\bibitem[Silich et al. (2010)]{Silich10} Silich, S., Tenorio-Tagle, G., Mu{\~n}oz-Tu{\~n}{\'o}n, C., Hueyotl-Zahuantitla, F., W{\"u}nsch, R., \& Palou{\v s}, J.,\ \apj, 711, 25

\bibitem[Steidel et al.(1999)]{Steidel99} Steidel, C.~C., 
Adelberger, K.~L., Giavalisco, M., Dickinson, M., 
\& Pettini, M.\ 1999, \apj, 519, 1 

\bibitem[Storey \& Hummer(1995)]{Storey95} Storey, P.~J., \& 
Hummer, D.~G.\ 1995, MNRAS, 272, 41 

\bibitem[Tenorio-Tagle et al.(2006)]{GTT06} Tenorio-Tagle, 
G., Mu{\~n}oz-Tu{\~n}{\'o}n, C., P{\'e}rez, E., Silich, S., 
\& Telles, E.\ 2006, \apj, 643, 186 

\bibitem[Telles et al.(1997)]{TMT97} Telles, E., Melnick, J., \& Terlevich, R.\ 1997, \mnras, 288, 78 

\bibitem[Terlevich et al.(1991)]{Terlevich91} Terlevich, R., Melnick, J., Masegosa, J., Moles, M., \& Copetti, M.~V.~F.\ 1991, \aaps, 91, 285 

\bibitem[Vacca \& Conti(1992)]{Vacca92} Vacca, W.~D., \& Conti, P.~S.\ 1992, \apj, 401, 543 

\bibitem[Vaduvescu et al.(2006)]{Vaduvescu06} Vaduvescu, O., 
Richer, M.~G., \& McCall, M.~L.\ 2006, \aj, 131, 1318 

\bibitem[van der Wel et al.(2011)]{vanderWel11} van der Wel, A., 
Straughn, A.~N., Rix, H.-W., et al.\ 2011, \apj, 742, 111 

\bibitem[van Zee et al.(1998)]{Vanzee98} van Zee, L., Salzer, J.~J., \& Haynes, M.~P.\ 1998, \apjl, 497, L1 

\bibitem[van Zee \& Haynes(2006)]{Vanzee06} van Zee, L., \& Haynes, M.~P.\ 2006, \apj, 636, 214 

\bibitem[Vila Costas \& Edmunds(1993)]{VilaCostas93} Vila Costas, M.~B., \& Edmunds, M.~G.\ 1993, \mnras, 265, 199 

\bibitem[Werk et al.(2004)]{Werk04} Werk, J.~K., Jangren, A., \& Salzer, J.~J.\ 2004, \apj, 617, 1004 

\end{thebibliography}
\end{document}